\documentclass[10pt,journal]{IEEEtran}
\IEEEoverridecommandlockouts 
\usepackage{blkarray}                                      
\usepackage{algpseudocode}                                 
\usepackage{algorithm}
\usepackage{graphicx}                                      
\usepackage{amsmath}
\usepackage{amssymb}
\usepackage{amsfonts}
\usepackage{amsthm}
\usepackage[mathcal]{eucal}
\usepackage{mathrsfs}
\usepackage{booktabs}
\usepackage{enumerate}
\usepackage{multirow}
\usepackage{subcaption}
\usepackage{color}
\usepackage{cite}                                          
\usepackage{comment}                                       
\usepackage{soul}                                          
\soulregister\cite7
\soulregister\ref7
\soulregister\pageref7
\usepackage{etoolbox}                                      
\usepackage{url}
\usepackage{nth}                                           
\usepackage{bm}                                            
\usepackage{courier}
\usepackage{balance}
\usepackage{threeparttable}
\usepackage{xcolor,colortbl}
\usepackage{footnote}
\usepackage{listings}
\usepackage{setspace}                                      

\usepackage{verbatim}
\usepackage[bookmarks=false]{hyperref}
\hypersetup{
    colorlinks = true,
    citecolor  = blue,
    linkcolor  = blue,
    urlcolor   = blue,
}
\usepackage{tikz}
\usetikzlibrary{patterns,snakes}
\usetikzlibrary{positioning,calc,fit,decorations.pathmorphing,shapes.geometric, shapes.gates.logic.US, calc}
\usetikzlibrary{arrows,arrows.meta,decorations.markings,shapes,shapes.arrows}
\usetikzlibrary{decorations,decorations.pathreplacing}
\usetikzlibrary{backgrounds}
\usepackage{filecontents}                                  
\usepackage{pgfplots}
\usepackage{pgfplotstable}
\usepackage{scalefnt}
\pgfplotsset{compat=newest}
\usepackage{caption}
\usepackage{cleveref}
\Crefformat{figure}{Fig.~#2#1#3}                           
\Crefname{subfigure}{Fig.}{Figs.}
\Crefname{figure}{Fig.}{Figs.}
\Crefformat{table}{TABLE~#2#1#3}                           
\usepackage[figuresright]{rotating}

\definecolor{CUHKorange}{RGB}{244,106,18} 
\definecolor{CUHKblue}{RGB}{0,111,190}    
\definecolor{CUHKgreen}{RGB}{0,127,128}   
\definecolor{CUHKred}{RGB}{228,46,36}     
\definecolor{CUHKyellow}{RGB}{198,148,34} 
\definecolor{CUHKdark}{RGB}{114,44,114}   
\definecolor{CUHKmiddle}{RGB}{144,44,144} 
\definecolor{CUHKlight}{RGB}{167,44,167} 
\definecolor{CUHKpurple}{RGB}{117,15,109}
\definecolor{CUHKgold}{RGB}{221,163,0}
\definecolor{CUHKribbon}{RGB}{244,223,176}
\definecolor{CUHKblack}{RGB}{34,24,21}

\renewcommand{\bf}[1]{\textbf{#1}}
\renewcommand{\tt}[1]{\texttt{#1}}

\newcommand{\minisection}[1]{\vspace{.1in}\noindent{\textbf{#1}}}

\newcommand{\vol}{\ensuremath{\mathrm{vol}}}

\usepackage{tcolorbox}
\tcbuselibrary{skins,breakable}
    {\endtcolorbox}
    {\endtcolorbox}

\usepackage[top=0.78in,bottom=0.86in,left=0.62in,right=0.62in]{geometry}
\newtheorem{mydefinition}{\textbf{Definition}}
\newtheorem{mytheorem}{\textbf{Theorem}}

\newtheorem{mylemma}{\textbf{Lemma}}

\crefname{mytheorem}{Theorem}{Theorems}
\crefname{mylemma}{Lemma}{Lemmas}
\crefname{myclaim}{Claim}{Claims}
\crefname{myproperty}{Property}{Properties}
\crefname{mycorollary}{Corollary}{Corollaries}

\algrenewcommand\textproc{\texttt}

\makeatletter
\let\OldStatex\Statex
\renewcommand{\Statex}[1][3]{%
  \setlength\@tempdima{\algorithmicindent}%
  \OldStatex\hskip\dimexpr#1\@tempdima\relax
}
\makeatother

\algblockdefx[Foreach]{Foreach}{EndForeach}[1]{\textbf{for each} #1 \textbf{do}}{\textbf{end for}}
\makeatletter
\ifthenelse{\equal{\ALG@noend}{t}}{\algtext*{EndForeach}}{}
\makeatother

\RequirePackage[normalem]{ulem} 
\RequirePackage{color}\definecolor{RED}{rgb}{1,0,0}\definecolor{BLUE}{rgb}{0,0,1} 

\graphicspath{{./figs/}{../}}

\newcommand{\ubold}{\fontseries{b}\selectfont}
\robustify\ubold

\newcommand{\power}[2][10]{#1\textsuperscript{#2}}

\usepackage{calc}
\usepackage{mathtools}
\usepackage{siunitx}
\usepackage{tikz}
\usetikzlibrary{positioning,calc,fit,decorations.pathmorphing,shapes.geometric,shapes.gates.logic.US,calc}
\usepackage{tikz-3dplot}
\usetikzlibrary{backgrounds,arrows.meta,shapes,shadows}
\usepackage{dsfont}
\definecolor{myred}{HTML}{ee8581}
\definecolor{myblue}{HTML}{40739e}
\definecolor{mypurple}{HTML}{7f7ff7}
\definecolor{mygreen}{HTML}{87b795}
\definecolor{mygray}{HTML}{d3d3d3}
\definecolor{mycyan}{HTML}{90aab0}
\definecolor{mybrown}{HTML}{a87b7b}
\definecolor{mydeepblue}{HTML}{40739e}
\definecolor{mylightblue}{HTML}{b0e0e6}
\definecolor{mydeepred}{HTML}{eb6f6b}
\definecolor{bistre}{HTML}{3d2b1f}
\definecolor{bole}{HTML}{79443b}
\definecolor{slategray}{RGB}{112,128,144}
\definecolor{darkgray}{RGB}{47,79,79}
\definecolor{chazblue}{RGB}{64,115,159}
\definecolor{chazred}{RGB}{192,143,143}

\definecolor{slategray}{RGB}{112,128,144}
\definecolor{darkgray}{RGB}{47,79,79}
\definecolor{rosybrown}{RGB}{188,143,143}
\definecolor{deepblue}{RGB}{64,115,158}

\crefname{mydefinition}{Definition}{Definitions}
\crefname{mytheorem}{Theorem}{Theorems}
\crefname{mylemma}{Lemma}{Lemmas}

\makeatletter
\newcommand{\labeltext}[3][]{%
  \@bsphack%
  \csname phantomsection\endcsname
  \def\tst{#1}%
  \def\labelmarkup{}
  \def\refmarkup{}%
  \ifx\tst\empty\def\@currentlabel{\refmarkup{#2}}{\label{#3}}%
  \else\def\@currentlabel{\refmarkup{#1}}{\label{#3}}\fi%
  \@esphack%
  \labelmarkup{#2}
}
\makeatother

\begin{document}

\title{
  Analytical Heterogeneous Die-to-Die 3D Placement with Macros
}

\author
{
  Yuxuan Zhao\textsuperscript{$\dag$}, \quad
  Peiyu Liao\textsuperscript{$\dag$},  \quad
  Siting Liu,  \quad
  Jiaxi Jiang, \quad
  Yibo Lin,    \quad
  Bei Yu\\
  \thanks{\textsuperscript{$\dag$}These authors contributed equally.}
  \thanks{
    This work is partially supported by
    The Research Grants Council of Hong Kong SAR (No.~CUHK14210723 and No.~CUHK14211824),
    and AI Chip Center for Emerging Smart Systems (ACCESS), Hong Kong.
    (\textit{Corresponding authors: Bei Yu})
  }
  \thanks{Yuxuan Zhao, Peiyu Liao, Siting Liu, Jiaxi Jiang and Bei Yu are with the Department of Computer Science and Engineering, The Chinese University of Hong Kong, NT, Hong Kong SAR.}
  \thanks{Y.~Lin is with the School of Integrated Circuits, Peking University, China, Institute of Electronic Design Automation, Peking University, Wuxi, China, and Beijing Advanced Innovation Center for Integrated Circuits, Beijing, China}
}

\maketitle
\pagestyle{plain}

\begin{abstract}
This paper presents an innovative approach to 3D mixed-size placement in heterogeneous face-to-face (F2F) bonded 3D ICs. 
We propose an analytical framework that utilizes a dedicated density model and a bistratal wirelength model, effectively handling macros and standard cells in a 3D solution space.
A novel 3D preconditioner is developed to resolve the topological and physical gap between macros and standard cells.
Additionally, we propose a mixed-integer linear programming (MILP) formulation for macro rotation to optimize wirelength.
Our framework is implemented with full-scale GPU acceleration, leveraging an adaptive 3D density accumulation algorithm and an incremental wirelength gradient algorithm. 
Experimental results on ICCAD 2023 contest benchmarks demonstrate that our framework can achieve {5.9}\% quality score improvement compared to the first-place winner with {4.0}$\times$ runtime speedup.
Additional experiments on modern RISC-V designs further validate the generalizability and superiority of our framework.
\end{abstract}

\section{Introduction}
\label{sec:intro}
\IEEEPARstart{A}{s} technology scaling approaches its physical limits, 3D integrated circuits (3D ICs) have emerged as a viable solution for extending Moore's Law.
By stacking multiple dies vertically, 3D ICs can integrate devices such as CMOS, SRAM, and RRAM with one or multiple technology nodes onto a single chip~\cite{DATE21-Sung}. 
However, circuit components like memory and analog blocks become the bottleneck of integration, which tend to scale at a slower pace than their logic counterpart.
Heterogeneous 3D ICs can benefit by using advanced technology nodes for standard cells without worrying about the technology node of the hard IPs, achieving better performance, area, and cost. 
Intel's Meteor Lake~\cite{HC2022-Meteor} serves as a notable example of such technology adoption. 

There are three main variants of 3D ICs: through-silicon-via (TSV) based, monolithic, and face-to-face (F2F) bonding.
The large pitches and parasitics of TSVs~\cite{dong2010fabrication} restrict TSV-based 3D ICs to few inter-die connections, thereby limiting the performance benefits.
While monolithic 3D (M3D) integration enables fine-grained vertical interconnects~\cite{samal2016monolithic,ICCAD20-Pentapati}, the manufacturing yield is low due to the sophisticated process steps.
F2F bonded 3D ICs~\cite{song2015coupling,jung2014enhancing} are made up of two prefabricated dies connected via hybrid bonding terminals (HBTs) on the top-most metal layer.
The ease of manufacturing and the small size of bonding terminals enable a high integration density at a low cost, making it a preferred approach~\cite{ISPD18-Ku,DATE20-Bamberg}.

3D placement remains a challenging problem in the physical design flow of 3D ICs.
Existing methodologies are either designed for standard cell 3D placement or aim to handle macros and standard cells together in mixed-size designs.
Recent placers~\cite{ICCAD23-Zhao,DAC23-Chen,TCAD23-Liao} for F2F bonded 3D ICs focus on standard cell placement~\cite{ICCAD22-Hu}.
iPL-3D~\cite{ICCAD23-Zhao} models the problem using bilevel programming to optimize partitioning and placement alternatively.
To model the heterogeneous integration, MTWA~\cite{DAC23-Chen} uses a sigmoid-based pin transition function, and the bistratal wirelength model~\cite{TCAD23-Liao} proposes the finite difference approximation for accurate wirelength modeling.
However, with memory-intensive applications such as machine learning proliferating, 
numerous memory macros are integrated into modern processors and accelerators to enhance performance.
A 3D placer capable of handling both standard cells and macros is essential to obtain the expected benefits~\cite{DATE20-Bamberg} for these mixed-size designs.

Existing 3D mixed-size placers form two broad categories: pseudo-3D and true-3D.
Pseudo-3D placers~\cite{ICCAD16-Chang,ICCAD20-Pentapati,vanna2021snap,TCAD17-Panth, ISPD18-Ku} separate the partitioning and placement phases, and adopt 2D placement tools to determine instance locations.
{Cascade2D}~\cite{ICCAD16-Chang} implements an M3D design using the partitioning-first flow.
To fully utilize the physical information, recent partitioning-last flows~\cite{TCAD17-Panth, ISPD18-Ku} perform tier partitioning after an intermediate placement stage. 
These design flows introduce placement blockages to consider pre-placed macros from the floorplan stage. 
However, pseudo-3D placers cannot fully explore the overall solution space, and their performance is particularly sensitive to partitioning results, exacerbated by the presence of macros.

Differently, true-3D placers~\cite{TCAD13-Luo, TCAD13-Hsu, ISPD16-Lu} relax the discrete tier partitioning and adopt analytical approaches.
The analytical placers perform mixed-size placement in a 3D cuboid region based on the smoothed wirelength model and density model.
{NTUPlace3-3D}~\cite{TCAD13-Hsu} utilizes a bell-shaped density model considering TSV insertion.
The state-of-the-art (SOTA) analytical 3D placer, ePlace-3D~\cite{ISPD16-Lu}, models the density constraint as a 3D electrostatic field.
Despite their efficiency in handling macros and standard cells, existing true-3D placers focus on TSV minimization without an accurate model for heterogeneous integration.

In summary, the aforementioned previous approaches are hardly applicable to mixed-size designs in heterogeneous F2F bonded 3D ICs.
Most pseudo-3D placers~\cite{ICCAD16-Chang, TCAD17-Panth} rely on the FM min-cut partitioning algorithm~\cite{fiduccia1988linear} and fail to utilize the advantages of F2F bonding technology.
Conventional true-3D placers~\cite{TCAD13-Luo, TCAD13-Hsu, ISPD16-Lu} do not support heterogeneous integration and employ a simplistic 3D net bounding box wirelength model, neglecting the wirelength reduction through inter-die connections.
Although recent studies~\cite{DAC23-Chen,TCAD23-Liao} have improved wirelength models to better accommodate heterogeneous technology nodes, the placers lack key innovations for the significant topological and physical difference between macros and standard cells, resulting in challenges with optimization convergence.

{
GPU acceleration has achieved great success in 2D placement~\cite{DAC19-Lin,DAC22-Liu}.
Liao \textit{et al.}~\cite{TCAD23-Liao} pioneered GPU acceleration for 3D placement, but their acceleration techniques are limited to standard cell placement, resulting in significant load balancing issues in mixed-size scenarios.
In addition, the approach they employed for bistratal wirelength model~\cite{TCAD23-Liao}  is hampered by high computational complexity.
Innovations are needed for efficient 3D mixed-size placement on GPU.
}

In this paper, we propose an analytical approach to 3D mixed-size placement in heterogeneous F2F bonded 3D ICs.
Leveraging a dedicated density model and a bistratal wirelength model, our framework effectively optimizes instance partitioning and locations in a 3D solution space.
Our contributions are summarized as follows.
\begin{itemize}
    \item We propose an analytical 3D mixed-size placement framework with a density model and a bistratal wirelength model, incorporating a novel 3D preconditioner, for heterogeneous F2F bonded 3D ICs.
    \item A mixed-integer linear programming (MILP) formulation is proposed to assign macro rotations for wirelength optimization.
    \item We implement our framework with full-scale GPU acceleration, leveraging adaptive 3D density accumulation and incremental wirelength gradient algorithms.
    \item Experimental results on ICCAD 2023 contest benchmarks demonstrate that our framework can achieve {5.9}\% quality score improvement over the first-place winner with {4.0}$\times$ runtime speedup. 
    \item {We also evaluated our framework on modern RISC-V designs. Compared to the baseline, our placer achieves {20}\% better wirelength with {12.0}$\times$ runtime speedup.}
\end{itemize}

The remainder of this paper is organized as follows.
\Cref{sec:prelim} provides the background and the problem formulation.
\Cref{sec:overall-framework} presents the overall mixed-size placement flow of the proposed framework for heterogeneous F2F bonded 3D ICs.
In~\Cref{sec:agl}, we detail our density and wirelength algorithms.
\Cref{sec:results} presents experimental results and related analysis, followed by conclusion in~\Cref{sec:conclusion}.

\section{Preliminaries}
\label{sec:prelim}

\subsection{3D Analytical Global Placement}
\label{subsec:prelim-analytical-place}
Given a netlist $(V,E)$ where $V=\{v_1,\cdots,v_n\}$ is the instance set and $E=\{e_1,\cdots,e_m\}$ is the net set, all the instances are placed within a 3D cuboid region $\Omega = [0, d_{x}] \times [0, d_{y}] \times [0, d_{z}]$.
And we use $V_{M} \subset V$ and $E_{M} \subset E$ to denote the movable macros and the nets connecting the macros.
Let $\bm{v} = (\bm{x}, \bm{y}, \bm{z})$ denote the physical coordinates of the instances.
The placement objective is to minimize the total half-perimeter wirelength (HPWL) while satisfying the target density constraints.
Conventionally, the 3D HPWL is adopted as the objective function defined below.

\begin{mydefinition}[3D HPWL]
  \label{def:3d-hpwl}
  Given instance locations $\bm{v} = (\bm{x},\bm{y},\bm{z})$, the 3D HPWL of any net $e\in E$ is given by
  \begin{equation}
    W_e(\bm{v})=p_{e}(\bm{x})+p_{e}(\bm{y})+\alpha \cdot p_{e}(\bm{z}),
    \label{eq:3d-hpwl-def}
  \end{equation}
  where $p_e(\bm{u})=\max_{v_i\in e}u_i - \min_{v_i\in e}u_i$ denotes the partial HPWL along one axis, and a weight factor $\alpha\geq0$ is introduced for the vertical interconnects in 3D ICs.
\end{mydefinition}
To model the density constraints, the cuboid region $\Omega$ is uniformly divided into $N_x \times N_y \times N_z$ bins denoted as set $B$.
And the density $\rho_{b}$ in each bin should not exceed the target density $\rho_t$. 
The nonlinear placement optimization is formulated as 
\begin{equation}
  \displaystyle\min_{\bm{v}} \displaystyle\sum_{e\in E}W_e(\bm{v}) \quad
  \mathrm{s.t.} \  \rho_{b}(\bm{v}) \leq \rho_{t}, \forall b \in B.
  \label{eq:nonlinear-place}
\end{equation}

Analytical methods conduct the 3D global placement using gradient-based optimization.
As $p_e(\cdot)$ in 3D HPWL is nonsmooth and nonconvex, it is approximated by a differentiable wirelength model, \emph{e.g.}, the weighted-average model~\cite{TCAD13-Hsu} given a smoothing parameter $\gamma>0$,
\begin{equation}
  \hat{p}_{e}(\bm{x})=\frac{\sum_{v_i\in e}x_i\mathrm{e}^{\frac{1}{\gamma}x_i}}{\sum_{v_i\in e}\mathrm{e}^{\frac{1}{\gamma}x_i}}-\frac{\sum_{v_i\in e}x_i\mathrm{e}^{-\frac{1}{\gamma}x_i}}{\sum_{v_i\in e}\mathrm{e}^{-\frac{1}{\gamma}x_i}}.
  \label{eq:wa-model-approx}
\end{equation}
Similarly, a density model $U(\cdot)$ relaxes all the $|B|$ constraints in~\Cref{eq:nonlinear-place} and evaluates the overall density penalty within the entire region $\Omega$.
The state-of-the-art density model $U(\cdot)$ is the {eDensity} family~\cite{TODAES2015-Lu-ePlace,PLACE-TCAD2015-Lu, ISPD16-Lu} based on electrostatics field, converting instances $v_i\in V$ to charges.
The electric force spreads charges towards the equilibrium state, producing a globally even density distribution.
Putting the density penalty into the wirelength objective, the 3D analytical global placement is formulated as the following unconstrained optimization
\begin{equation}
  \min_{\bm{v}}\sum_{e\in E} \hat{W}_e(\bm{v})+\lambda \hat{U}(\bm{v}),
  \label{eq:analytical-placement}
\end{equation}
where $\hat{W}_e(\cdot)$ is the smoothed wirelength model, $\hat{U}(\cdot)$ is the smoothed density model, 
and $\lambda$ is the density weight introduced as the Lagrangian multiplier of the density constraints.

\subsection{Problem Formulation}
\label{subsec:problem-formulation}

This paper considers the 3D mixed-size placement problem specified in the ICCAD 2023 contest~\cite{ICCAD23-Hu}.
We intend to determine the locations of standard cells and macros on the two dies with the same or different technology nodes, and insert hybrid bonding terminals (HBTs) for die-to-die (D2D) vertical connections 
so that the total D2D wirelength and HBT cost are minimized while the following constraints are satisfied:
\begin{enumerate}
\item All the instances must be non-overlapping, and the standard cells must be aligned to rows and sites. HBT spacing constraints must be satisfied.
\item All the instances are placed on either top or bottom die, and the maximum utilization of each die must be satisfied.
\item For any crossing-die net, one and only one HBT is inserted for vertical connection.
\item All the standard cells cannot be rotated or mirrored. Macros, on the other hand, can be rotated with 0$^{\circ}$, 90$^{\circ}$, 180$^{\circ}$, and 270$^{\circ}$ counterclockwise without mirroring.
\end{enumerate}

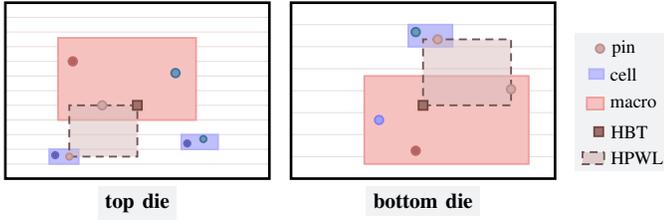
\begin{figure}[tb!]
  \centering\footnotesize
\begin{tikzpicture}[
    scale=.39,
    frame/.style={draw=rosybrown!60!black},
    base/.style={line width=.7pt},
    tmac/.style={base,fill=myred!50,draw=myred!80},
    tcell/.style={base,fill=mypurple!50,draw=mypurple!60},
    pin/.style={circle,inner sep=.4mm,line width=.7pt},
    hbt/.style={pin,rectangle,inner sep=.6mm,fill=bole!75,draw=bole},
    n1/.style={pin,fill=rosybrown!75,draw=rosybrown},
    n2/.style={pin,fill=myred!75!black,draw=myred!80!black},
    n3/.style={pin,fill=mypurple!75!black,draw=mypurple!80!black},
    n4/.style={pin,fill=deepblue!75,draw=deepblue}
  ]
  \pgfdeclarelayer{mid}
  \pgfdeclarelayer{fg}
  \pgfsetlayers{background,mid,main,fg}
  \draw[line width=1pt] (0,1.5) rectangle (9,7.5);
  \begin{pgfonlayer}{mid}
    \foreach\y in {4,...,14}{
      \draw[rosybrown!30] (0,.5*\y) -- (9,.5*\y);
    }
  \end{pgfonlayer}
  \coordinate (macro) at (1.8,3.5);
  \coordinate (cell) at (1.5,2);
  \coordinate (cell2) at (6.0,2.5);
  \filldraw[tmac] (macro) rectangle++ (4.7,2.8);
  \filldraw[tcell] (cell) rectangle++ (1.0,0.5);
  \filldraw[tcell] (cell2) rectangle++ (1.25,0.5);
  \node[n2] at ($(macro)+(0.5,2)$) {};
  \node[n4] at ($(macro)+(4.0,1.6)$) {};
  \node[n3,inner sep=.3mm] at ($(cell)+(0.2,.3)$) {};
  \begin{pgfonlayer}{fg}
    \node[n1] (pin1) at ($(macro)+(1.5,0.5)$) {};
    \node[n1,inner sep=.3mm] (pin2) at ($(cell)+(0.68,0.25)$) {};
    \node[hbt] (hbt1) at (4.5,4) {};
  \end{pgfonlayer}
  \node[n3,inner sep=.3mm] at ($(cell2)+(0.2,0.2)$) {};
  \node[n4,inner sep=.3mm] at ($(cell2)+(0.75,0.35)$) {};
  \node[fit={(pin1.center)(pin2.center)(hbt1.center)},inner sep=0,line width=.7pt,
    fill=rosybrown!60,fill opacity=.5,draw=rosybrown!60!black,densely dashed] {};
  \node[anchor=north,fill=slategray!10] at (4.5,1.25) {\textbf{top die}};
  \node[inner sep=0] at (4.5,0.1) {};
\end{tikzpicture}
\hspace{.2em}
\begin{tikzpicture}[
    scale=.39,node distance=3mm,
    frame/.style={draw=rosybrown!60!black},
    base/.style={line width=.7pt},
    tmac/.style={base,fill=myred!50,draw=myred!80},
    tcell/.style={base,fill=mypurple!50,draw=mypurple!60},
    pin/.style={circle,inner sep=.4mm,line width=.7pt},
    hbt/.style={pin,rectangle,inner sep=.6mm,fill=bole!75,draw=bole},
    n1/.style={pin,fill=rosybrown!75,draw=rosybrown},
    n2/.style={pin,fill=myred!75!black,draw=myred!80!black},
    n3/.style={pin,fill=mypurple!75,draw=mypurple},
    n4/.style={pin,fill=deepblue!75,draw=deepblue}
  ]
  \pgfdeclarelayer{mid}
  \pgfdeclarelayer{fg}
  \pgfsetlayers{background,mid,main,fg}
  \draw[line width=1pt] (0,1.5) rectangle (9,7.5);
  \begin{pgfonlayer}{mid}
    \foreach\y in {3,...,9}{
      \draw[rosybrown!30] (0,.75*\y) -- (9,.75*\y);
    }
  \end{pgfonlayer}
  \coordinate (macro) at (2.5,2);
  \coordinate (cell) at (4,6);
  \filldraw[tmac] (macro) rectangle++ (5.6,3);
  \filldraw[tcell] (cell) rectangle++ (1.5,.75);
  \node[n2] at ($(macro)+(1.75,0.45)$) {};
  \node[n3] at ($(macro)+(0.5,1.5)$) {};
  \node[n4] at ($(cell)+(0.25,.5)$) {};
  \begin{pgfonlayer}{fg}
    \node[n1] (pin3) at ($(macro)+(5,2.55)$) {};
    \node[n1] (pin4) at ($(cell)+(1.0,0.25)$) {};
    \node[hbt] (hbt1) at (4.5,4) {};
  \end{pgfonlayer}
  \node[fit={(pin3.center)(pin4.center)(hbt1.center)},inner sep=0,line width=.7pt,
    fill=rosybrown!60,fill opacity=.5,draw=rosybrown!60!black,densely dashed] {};
  \node[anchor=north,fill=slategray!10] at (4.5,1.25) {\textbf{bottom die}};
  \node[inner sep=0] at (4.5,0.1) {};
\end{tikzpicture}
\hspace{.2em}
\begin{tikzpicture}[
    scale=.32,node distance=2mm,
    frame/.style={draw=rosybrown!60!black},
    pin/.style={circle,inner sep=.4mm,line width=.7pt},
    hbt/.style={pin,rectangle,inner sep=.6mm,fill=bole!75,draw=bole},
    n1/.style={pin,fill=rosybrown!75,draw=rosybrown},
  ]
  \scriptsize
  \node[n1,anchor=north west] (pin label) {};
  \node[anchor=west] (pin text) at (pin label.east) {pin};
  \node[below=of pin label.south east,inner sep=0] (anchor) {};
  \draw[line width=.7pt,fill=mypurple!50,draw=mypurple!60]
  (anchor) rectangle++ (-.6,-.4);
  \node[inner sep=0] at ($(anchor)+(0,-.4)$) (std label) {};
  \node[anchor=west] (std text) at ($(anchor)+(0,-0.2)$) {cell};
  \node[below left=2mm and 0 of std label.south east,inner sep=0] (anchor) {};
  \draw[line width=.7pt,fill=myred!50, draw=myred!80]
  (anchor) rectangle++ (-.7,-.5) node[] (macro label) {};
  \node[anchor=west] (macro text) at ($(anchor)+(0,-0.25)$) {macro};
  \coordinate[below=of macro label.south east] (anchor);
  \node[hbt] (hbt label) at ($(anchor)+(0.2,0)$) {};
  \node[anchor=west] (hbt text) at (hbt label.east) {HBT};
  \coordinate[below=of hbt label.south east] (anchor);
  \draw[line width=.7pt,fill=rosybrown!60,fill opacity=.7,draw=rosybrown!60!black,densely dashed]
  (anchor) rectangle++ (-.9,-.5) node[] (hpwl label) {};
  \node[anchor=west] (hpwl text) at ($(anchor)+(0,-0.25)$) {HPWL};
  \begin{pgfonlayer}{background}
    \node[fill=slategray!10,inner sep=0,
      fit={(pin label)(pin text)(hpwl label)(hpwl text)}] (legend) {};
  \end{pgfonlayer}
  \node[below=2em of hpwl text] {};
\end{tikzpicture}
\vskip-.25em%
  \caption{D2D wirelength of a net is the sum of the wirelength of the top net and bottom net. HBTs are on the top-most layer for both dies. Pins connected by a net are in the same color.}
  \label{fig:d2d-wl}
\end{figure}

It is worth noting that the cells and macros may be fabricated using different technology nodes on different dies, \emph{i.e.}, the instance height, width, and pin location would be different.
And the center points of the HBTs are included in the wirelength calculation for accurate modeling of F2F bonded ICs, as illustrated in~\Cref{fig:d2d-wl}. Therefore, the instance partition $\bm{\delta}$ must be explicitly considered.

A partition is determined by a binary vector $\bm{\delta}\in \{0,1\}^n$, where $\delta_i=0$ indicates that $v_i\in V$ is placed on the bottom die, otherwise on the top die.
$\delta_i$ can be derived from the instance center $z$-coordinate by $\delta_i=\mathds{1}_{\mathbb{R}^+}(z_i - \frac{d_z}{2})$, where $\mathds{1}_{\mathbb{R}^+}(\cdot)$ is the indicator function of positive real numbers.
Accordingly, the HBT $t_e$ is inserted for crossing-die net $e$ with $C_{e}(\bm{\delta}) = \max_{v_i\in e}\delta_i - \min_{v_i\in e}\delta_i = 1$, which means the net $e$ connects the instances placed on the different dies.
The die-to-die (D2D) wirelength~\cite{ICCAD23-Hu} includes the top net $\hat{e}^{+} = e^{+}\cup\{t_e\}$ and the bottom net $\hat{e}^{-} = e^{-}\cup\{t_e\}$ considering both instances and HBTs, where $e^{+} = \{v_i \in e : \delta_i = 1\}$ and $e^{-} = \{v_i \in e : \delta_i = 0\}$.

\begin{mydefinition}[D2D HPWL]
    \label{def:net-wirelength}
    Given partition $\bm{\delta}$, the die-to-die (D2D) HPWL of net $e$ is defined by $W_{e}=W_{\hat{e}^{+}}+W_{\hat{e}^{-}}$, where
    \begin{equation}
      \begin{aligned}
        W_{\hat{e}^{+}}&=p_{\hat{e}^{+}}(\bm{x})+p_{\hat{e}^{+}}(\bm{y}),\\
        W_{\hat{e}^{-}}&=p_{\hat{e}^{-}}(\bm{x})+p_{\hat{e}^{-}}(\bm{y}).
      \end{aligned}
      \label{eq:exact-wirelength}
    \end{equation}
    If $C_{e}(\bm{\delta})=0$, it reduces to the 2D net HPWL without the HBT.
\end{mydefinition}

Based on~\Cref{def:net-wirelength}, we formally define the 3D die-to-die placement problem as follows.
\begin{equation}
  \begin{array}{cll}
    \displaystyle\min_{\bm{x},\bm{y},\bm{z}}&\multicolumn{2}{l}{\displaystyle\sum_{e\in E}W_e(\bm{x},\bm{y},\bm{z}) + \beta\sum_{e\in E}C_{e}(\bm{\delta})}\\
    \mathrm{s.t.}& \rho_{b}(\bm{x},\bm{y},\bm{z}) \leq \rho_{t}, &\forall b \in B,\\
    &\delta_i=\mathds{1}_{\mathbb{R}^+}(z_i - \frac{d_z}{2}), &\forall v_i \in V,\\
    & \theta_i \in \{0^{\circ}, 90^{\circ}, 180^{\circ}, 270^{\circ}\}, &\forall v_i \in V_M,\\
    &\text{legality constraints,}&
  \end{array}
  \label{eq:high-level-formulation}
\end{equation}
where $W_e(\cdot)$ is the D2D HPWL, and $C_{e}(\bm{\delta})$ is the crossing-die net indicator. $\beta$ denotes the cost of each HBT provided by the design specification, and $\theta_i$ denotes the rotation of each macro. 
Following the 3D analytical approaches, we transform the above problem into unconstrained optimization in~\Cref{eq:analytical-placement}.
We adopt the bistratal wirelength model~\cite{TCAD23-Liao} and eDensity3D model~\cite{ISPD16-Lu}, respectively.
Dedicated customizations are proposed for accurate modeling of heterogeneous mixed-size designs, and full GPU acceleration is implemented in our framework for ultrafast performance.



\section{Proposed 3D Placement Framework}
\label{sec:overall-framework}

\iffalse
\begin{figure}[tb!]
  \centering
  \includegraphics[width=\linewidth]{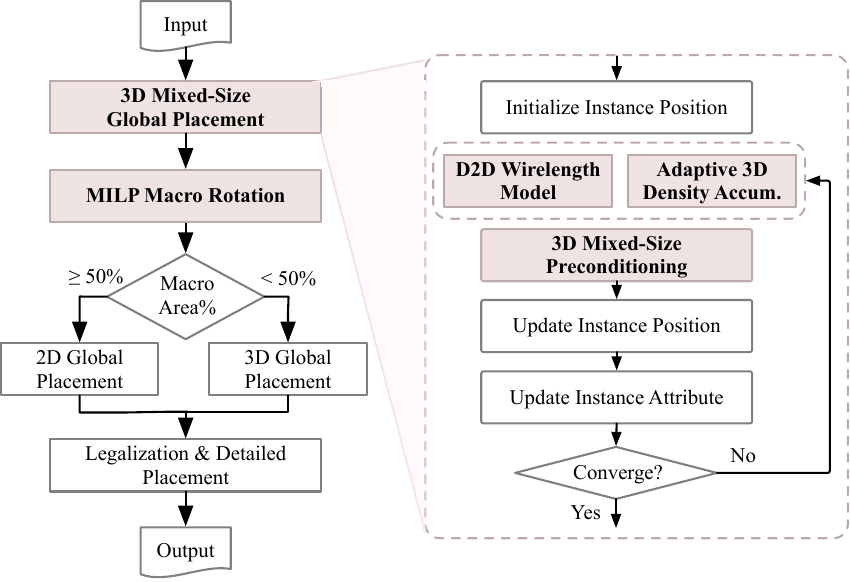}
  \caption{The overall 3D mixed-size placement flow.}
  \label{fig:placement-flow}
\end{figure}
\else
\begin{figure}[t]
  \centering
  \resizebox{\linewidth}{!}{
\begin{tikzpicture}[
    base/.style={draw=black},
    box/.style={rectangle,fill=none,
      inner sep=1.5mm,minimum width=5cm,line width=1pt,align=center},
    innerbox/.style={box,base},
    dash color/.style={draw=rosybrown!60!black},
    colored/.style={fill=rosybrown!40,dash color},
    stepbox/.style={innerbox,draw=rosybrown!60!black,
      densely dashed,fill=none,inner sep=2mm},
    seg/.style={base,line width=1pt},
    card/.style={align=center,base,
      line width=1pt,fill=white,
      inner sep=2mm,node distance=.35cm,
      minimum width=4.0cm},
    freecard/.style={card,minimum width=1cm},
    document/.style={line width=1pt,shape=tape,base,
      fill=white,minimum width=2cm,tape bend top=none},
    multidocument/.style={doc,double copy shadow},
    lseg/.style={seg,line width=1pt,
      -{Triangle[width=5pt,length=5pt]}},
    node distance=3mm,
  ]
  \pgfdeclarelayer{mid1}
  \pgfdeclarelayer{mid2}
  \pgfdeclarelayer{mid3}
  \pgfsetlayers{background,mid3,mid2,mid1,main}
  \node[innerbox] (init) {Initialize Instance Locations};
  \coordinate[above=2mm of init] (beforeinit);
  \coordinate[below=1.15cm of init.south] (grads-anchor);
  \coordinate[left=0mm of grads-anchor] (grads);
  \node[innerbox,fill=rosybrown!15,left=.5mm of grads,minimum width=2cm,] (density-grad) {%
    \textbf{Density Gradient}\\\small{\textbf{(Adaptive Accum.)}}};
  \node[innerbox,fill=rosybrown!15,right=.5mm of grads,minimum width=2cm,minimum height=1cm] (wl-grad) {%
    \textbf{WL Gradient}\\\small{\textbf{(Incremental Comp.)}}};
  \node[box,fill=none,base,densely dashed,dash color,rounded corners=3pt,
    fit={(density-grad)(wl-grad)}] (all-grad) {};

  \path let\p1=(all-grad.north east),\p2=(wl-grad.north east) in
  node[anchor=south east] at (\x2,\y1) {\small \Cref{subsec:wl-acl}};
  \path let\p1=(all-grad.north west),\p2=(density-grad.north west) in
  node[anchor=south west] at (\x2,\y1) {\small\Cref{subsec:density-accu}};

  \draw[seg,-{Triangle[width=5pt,length=5pt]}]
  let\p1=(init.south),\p2=(all-grad.north) in (\x1,\y1)--(\x1,\y2);
  \node[innerbox,fill=rosybrown!25,below=1.15cm of grads-anchor] (3d-precond) {%
    \textbf{3D Mixed-Size}\\\textbf{Preconditioning}};
  \draw[seg,-{Triangle[width=5pt,length=5pt]}]
  let\p1=(all-grad.south),\p2=(3d-precond.north) in (\x2,\y1)--(\x2,\y2);
  \node[innerbox,fill=rosybrown!35, below=of 3d-precond] (update-pos) {Update Instance Locations};
  \node[innerbox,fill=rosybrown!40, below=of update-pos] (update-attr) {Update Instance Attributes};
  \draw[seg,-{Triangle[width=5pt,length=5pt]}] (3d-precond)--(update-pos);
  \draw[seg,-{Triangle[width=5pt,length=5pt]}] (update-pos)--(update-attr);
  \node[base,diamond,inner sep=1mm,align=center,line width=1pt,fill=rosybrown!45,
    below=2mm of update-attr,aspect=2] (check) {Converge?};
  \node[below=2mm of check] (aftercheck) {};
  \coordinate (grad-east) at ($(all-grad.east)+(.3,0)$);
  \path let\p1=(grad-east),\p2=(check) in coordinate (mid-point)
  at ($(\x1,\y1)!.5!(\x1,\y2)$);
  \draw[seg,-{Triangle[width=5pt,length=5pt]}](mid-point)|-(all-grad);
  \node[label=90:{No}] at (check.east) {};
  \node[label=-30:{Yes}] at (check.south) {};
  \begin{pgfonlayer}{mid2}
    \node[stepbox,rounded corners=3pt,fill=white,
      fit={(beforeinit)(all-grad)(aftercheck)(mid-point)}]
    (global-place) {};
  \end{pgfonlayer}
  \path let\p1=(check),\p2=(global-place.south) in
  coordinate (gp-end) at ($(\x1,\y2)+(0,.1)$);
  \path let\p1=(init),\p2=(global-place.north) in
  coordinate (gp-beg) at ($(\x1,\y2)-(0,.1)$);
  \begin{pgfonlayer}{mid2}
    \draw[seg] (update-attr)--(check.center);
    \draw[seg] (check.center)-|(mid-point);
    \draw[seg,-{Triangle[width=5pt,length=5pt]},base] (check.center)--(gp-end);
  \end{pgfonlayer}
  \draw[seg,-{Triangle[width=5pt,length=5pt]}] (gp-beg)--(init);
  
  \node[card,fill=rosybrown!14,
    above left=2.2cm and 1cm of global-place.west] (gp-card) {%
    3D Mixed-Size\\Global Placement};
  \node[anchor=south west] at (gp-card.north west) {\small\Cref{subsec:wirelength-density-model}};
  \begin{pgfonlayer}{mid3}
    \fill[fill=rosybrown!15](gp-card.north east)
         {[rounded corners=3pt]--(global-place.north west)--
           (global-place.north east)--(global-place.south east)--
           (global-place.south west)
         }--(gp-card.south east)--cycle;
  \end{pgfonlayer}
  \node[document,above=6mm of gp-card] (input) {\textbf{Input}};
  \node[card,
    below=5mm of gp-card] (macro-rot) {{MILP Macro Rotation}};
  \node[anchor=south west] at (macro-rot.north west) {\small\Cref{subsec:macro-rotation}};
  \node[base,diamond,aspect=2,fill=white,
    inner sep=1mm,line width=1pt,below=of macro-rot,
    align=center] (macro-check) {Macro\\Area\%};
  \node[label=90:{$\geq$50\%}] at (macro-check.west) {};
  \node[label=90:{$<$50\%}] at (macro-check.east) {};
  \node[freecard,fill=rosybrown!14,below right=2mm and -2mm of macro-check] (3d-gp) {%
    3D Mixed-Size \\ Global Placement};
  \node[freecard,below left=2mm and -2mm of macro-check] (2d-gp) {%
    Multi-Die 2D \\ Global Placement};
  \node[anchor=south west,align=center] at ($(2d-gp.north west)+(0.2,0)$) {\small{Section}\\\ref{subsec:2d-gp}};
  \node[anchor=south east,align=center] at ($(3d-gp.north east)+(-0.3,0)$) {\small{Section}\\\ref{subsec:wirelength-density-model}};
  \path let \p1=(macro-check),\p2=(2d-gp.south east) in
  coordinate (center-2d-3d-gp) at (\x1,\y2);
  \coordinate[below=2mm of center-2d-3d-gp] (lgdp-marker);
  \node[card,below=6mm of lgdp-marker] (lgdp-card) {%
    Legalization \&\\Detailed Placement};
  \node[anchor=south west,align=center] at (lgdp-card.north west) {\small\Cref{subsec:legal}\;\;\;\;\ref{subsec:dp}};
  \node[document,below=of lgdp-card] (output) {\textbf{Output}};
  \begin{pgfonlayer}{mid2}
    \draw[lseg] (macro-check.center)-|(2d-gp);
    \draw[lseg] (macro-check.center)-|(3d-gp);
  \end{pgfonlayer}
  \begin{pgfonlayer}{mid1}
    \draw[lseg] (input.center)--(gp-card);
    \draw[lseg] (gp-card)--(macro-rot);
    \draw[lseg] (macro-rot)--(macro-check);
    \draw[seg,line width=1pt] (2d-gp)|-(lgdp-marker);
    \draw[seg,line width=1pt] (3d-gp)|-(lgdp-marker);
    \draw[lseg] (lgdp-marker)--(lgdp-card);
    \draw[lseg] (lgdp-card)--(output);
  \end{pgfonlayer}
\end{tikzpicture}
  }
  \caption{The overall 3D mixed-size placement flow.}
  \label{fig:overall-flow}
\end{figure}
\fi
The overall placement flow of our framework is illustrated in~\Cref{fig:overall-flow}, which consists of four stages. 
First, our framework performs 3D mixed-size global placement (\Cref{subsec:wirelength-density-model}) to optimize the instance partitioning and locations simultaneously with the initial macro orientation 0$^{\circ}$.
Second, we optimize the macro rotations based on the physical information of the initial 3D placement solution. 
We propose a mixed-integer linear programming (MILP) formulation (\Cref{subsec:macro-rotation}) to minimize the wirelength.
Then, we perform global placement again with the optimized macro rotations to further improve wirelength.
Our framework applies multi-die 2D global placement (\Cref{subsec:2d-gp}) for the designs of macro area ratio larger than 50\%, which avoids the large macro density obstacle and leads to better macro placement.
For the designs with macro area ratio smaller than 50\%, we perform 3D mixed-size global placement to explore the entire solution space for better wirelength.
{
The macro area ratio is calculated using the technology information of the top die, which has a smaller feature size.
}
At last, we apply die-by-die legalization and detailed placement to obtain the final placement result.

\subsection{3D Mixed-Size Global Placement}
\label{subsec:wirelength-density-model}

\begin{figure}
  \centering
  \newcommand\xmaxfigonly{12}
\newcommand\ymaxfigonly{10}
\begin{tikzpicture}[
    line join=round,label distance=-1mm,
    box/.style={opacity=.8},
    elliparc/.style args={#1:#2:#3}{insert path={++(#1:#3) arc (#1:#2:#3)}}
  ]
  \pgfdeclarelayer{fg}
  \pgfsetlayers{main,fg}
  \begin{axis}[
      grid=both,scale=.58,unit vector ratio=1 1 4,
      minor tick num=1,clip=false,
      axis line style={draw=none},
      view={20}{20},line width=.8pt,
      zmin=0,zmax=2,z buffer=sort,
      major tick length=0pt,minor tick length=0pt,
      xmin=0,xmax=\xmaxfigonly,
      ymin=0,ymax=\ymaxfigonly,
      xticklabels=\empty,
      yticklabels=\empty,
      zticklabels={,0, $\frac{d_z}{2}$,$d_z$},
      colormap={gray}{rgb=(0.25,0.45,0.62),rgb=(0.75,0.56,0.56)}
    ]
    \addplot3[data cs=cart,surf,domain=0:\xmaxfigonly,y domain=0:\ymaxfigonly,samples=2,opacity=0.25]{0.5};
    \node[label=0:{$\frac{d_z}{4}$}] at (\xmaxfigonly,\ymaxfigonly,0.5) {};
    \filldraw[box,fill=rosybrown!70,draw=rosybrown] (1,1,0.5)--(8.5,1,0.5)--(8.5,8.5,0.5)--(1,8.5,0.5)--cycle;
    \filldraw[box,fill=deepblue!70,draw=deepblue] (9.1,3,0.5)--(10,3,0.5)--(10,6,0.5)--(9.1,6,0.5)--cycle;

    \node[circle,draw=rosybrown,fill=white,inner sep=.3mm] at (8.5,5.8,0.5) {};
    \node[circle,draw=rosybrown,fill=white,inner sep=.3mm] at (5.2,1,0.5) {};
    \node[circle,draw=rosybrown,fill=white,inner sep=.3mm] at (1,5.2,0.5) {};

    \node[circle,fill=deepblue,inner sep=.3mm] at (10,4.8,0.5) {};
    \node[circle,fill=deepblue,inner sep=.3mm] at (9.4,3,0.5) {};

    \draw[canvas is xz plane at y=4.5,darkgray,
      opacity=.7,-latex,line width=1pt] (11,1) [elliparc=0:-90:2 and 0.5];
    
    \draw[dashed,opacity=.5] (1,1,1.5)--(1,1,0.5);
    \draw[dashed,opacity=.5] (6,1,1.5)--(8.5,1,0.5);
    \draw[dashed,opacity=.5] (6,6,1.5)--(8.5,8.5,0.5);
    \draw[dashed,opacity=.5] (1,6,1.5)--(1,8.5,0.5);

    \addplot3[data cs=cart,surf,domain=0:\xmaxfigonly,y domain=0:\ymaxfigonly,samples=2,opacity=0.25]{1.5};
    \node[label=0:{$\frac{3d_z}{4}$}] at (\xmaxfigonly,\ymaxfigonly,1.5) {};
    \filldraw[box,fill=myred!70,draw=myred] (1,1,1.5)--(6,1,1.5)--(6,6,1.5)--(1,6,1.5)--cycle;
    \filldraw[box,fill=mypurple!70,draw=mypurple] (9.1,3,1.5)--(9.7,3,1.5)--(9.7,5,1.5)--(9.1,5,1.5)--cycle;

    \draw[canvas is xz plane at y=4.5,darkgray,
      opacity=.7,latex-,line width=1pt] (11,1) [elliparc=90:0:2 and 0.5];

    \node[circle,draw=myred,fill=white,inner sep=.3mm] at (6,4.2,1.5) {};
    \node[circle,draw=myred,fill=white,inner sep=.3mm] at (3.8,1,1.5) {};
    \node[circle,draw=myred,fill=white,inner sep=.3mm] at (1,3.8,1.5) {};

    \node[circle,fill=mypurple,inner sep=.3mm] at (9.7,4.2,1.5) {};
    \node[circle,fill=mypurple,inner sep=.3mm] at (9.3,3,1.5) {};

    \begin{pgfonlayer}{fg}
      \draw[line width=.8pt] (0,0,0)--(0,0,2)--(0,\ymaxfigonly,2)--
      (\xmaxfigonly,\ymaxfigonly,2)--(\xmaxfigonly,\ymaxfigonly,0)--(\xmaxfigonly,0,0)--cycle;
    \end{pgfonlayer}
  \end{axis}
\end{tikzpicture}
\begin{tikzpicture}[
    scale=.7,
    box/.style={opacity=.8,line width=1pt},
    circ/.style={circle,inner sep=.2mm,line width=1pt}
  ]
  \centering\footnotesize
  \node at (0,-.5) {}; 
  \node (left bottom) at (0,0) {};
  \filldraw[box,fill=rosybrown!70,draw=rosybrown] (0,0) rectangle++ (.78,.48);
  \node[circ,fill=white,draw=rosybrown] at (0,.24) {};
  \node[circ,fill=white,draw=rosybrown] at (.52,0) {};
  \node[circ,fill=white,draw=rosybrown] at (.78,.32) {};
  \node[anchor=west] (macro bottom) at (.85,.24) {macro on the bottom};
  \filldraw[box,fill=myred!70,draw=myred] (.26,.7) rectangle++ (.52,.36);
  \node[circ,fill=white,draw=myred] at (.26,.86) {};
  \node[circ,fill=white,draw=myred] at (.61,.7) {};
  \node[circ,fill=white,draw=myred] at (.78,.91) {};
  \node[anchor=west] at (.85,.88) {macro on the top};
  \filldraw[box,fill=deepblue!70,draw=deepblue] (.51,1.4) rectangle++ (.27,.48);
  \node[circ,fill=deepblue,draw=deepblue] at (.6,1.4) {};
  \node[circ,fill=deepblue,draw=deepblue] at (.78,1.72) {};
  \node[anchor=west] at (.85,1.64) {cell on the bottom};
  \node (left top) at (.6,2.34) {};
  \filldraw[box,fill=mypurple!70,draw=mypurple] (.6,2.1) rectangle++ (.18,.36);
  \node[circ,fill=mypurple,draw=mypurple] at (.66,2.1) {};
  \node[circ,fill=mypurple,draw=mypurple] at (.78,2.34) {};
  \node[anchor=west] (cell top) at (.85,2.22) {cell on the top};
  \begin{pgfonlayer}{background}
    \node[fit={(macro bottom)(cell top)(left bottom)(left top)},
      fill=slategray!10] {};
  \end{pgfonlayer}
\end{tikzpicture}
\let\xmaxfigonly\undefined
\let\ymaxfigonly\undefined
  \caption{
    Our density model and wirelength model consider instance partitioning explicitly for accurate modeling of heterogeneous technology nodes.
    The instance attributes are updated dynamically in the global placement stage.
    The macro size transition is smoothed for stable density optimization.
  }
  \label{fig:node-3dplot}
\end{figure}
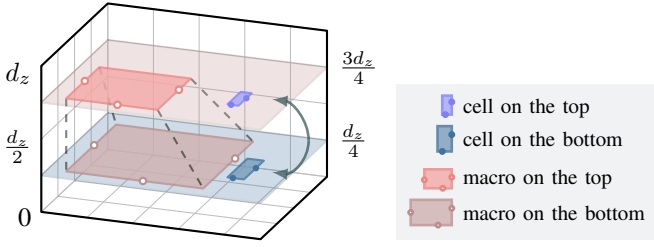
The heterogeneous F2F bonded ICs bring unique challenges for global placement.
The instance attributes including size and pin offsets are different on the two dies for heterogeneous technology nodes, and the macros show particularly large variation.
Such property requires our density model and wirelength model to consider the instance partitioning explicitly for accurate modeling.
\vskip .5em%

\minisection{Electrostatics-Based Density Model}.
The state-of-the-art eDensity3D~\cite{ISPD16-Lu} model sets the electric quantity $q_i$ as the physical volume of instance $v_i$.
To consider the heterogeneous technology nodes, we update the attributes of instance dynamically according to  the $z$-coordinate, \emph{i.e.}, $\delta_i=\mathds{1}_{\mathbb{R}^+}(z_i-\frac{d_z}{2})$, as shown in~\Cref{fig:node-3dplot}.
Let $w_i^{+}$ and $h_i^{+}$ denote the instance width and height on the top die, and $w_i^{-}$ and $h_i^{-}$ on the bottom die.
The dynamic width $w_i$ and height $h_i$ can be derived as
\begin{equation}
  \begin{aligned}
    w_i&=\delta_iw_i^{+}+(1-\delta_i)w_i^{-},\\
    h_i&=\delta_ih_i^{+}+(1-\delta_i)h_i^{-}.
  \end{aligned}
  \label{eq:std-size-tech}
\end{equation}
To accommodate the D2D placement, we set all instances with the same depth $d = \frac{1}{2}d_{z}$ so that the instances can be distributed to exactly two dies.
Although the update scheme provides accurate heterogeneous information, the step transition introduces discreteness for density optimization. 
The impact of standard cells is small, but the large variation in macro size incurs sudden changes in the density map, as shown in~\Cref{fig:node-3dplot}, resulting in challenges with convergence.
For any macro $v_i\in V_M$, we propose to linearly transform macro width and height as
\begin{equation}
  \begin{aligned}
    w_i&=\Big(\frac{2z_{i}}{d_{z}} - \frac{1}{2}\Big)w_i^{+}+\Big(\frac{3}{2} - \frac{2z_{i}}{d_{z}}\Big)w_i^{-},\\
    h_i&=\Big(\frac{2z_{i}}{d_{z}} - \frac{1}{2}\Big)\makebox[\widthof{$w_i^{+}$}][c]{$h_i^{+}$}+\Big(\frac{3}{2} - \frac{2z_{i}}{d_{z}}\Big)h_i^{-}.
  \end{aligned}
  \label{eq:macro-size-tech}
\end{equation}
The movable range of $z_i$ is $[\frac{d_z}{4}, \frac{3d_z}{4}]$ based on our depth setting.
While \cite{DAC23-Chen,DAC24-Chen} adopt nonlinear size transformation for both standard cells and macros, our approach only scales the macro size for more accurate heterogeneous modeling.

The eDensity3D models the density penalty $\hat{U}$ as the total potential energy of the system $\hat{U}(\bm{v}) = \sum_{v_i \in V}q_i\phi_i(\bm{v})$.
It computes the potential map $\phi(\bm{v})$ by solving the 3D Poisson's equation
\begin{equation}
  \begin{array}{rl}
    \Delta\phi(\bm{v})=-\rho(\bm{v}), & \bm{v}\in\Omega\\
    \hat{\bm{n}}\cdot\nabla\phi(\bm{v})=0, & \bm{v}\in\partial\Omega.\\
  \end{array}
  \label{eq:3d-poisson-equation}
\end{equation}
eDensity3D solves the Poisson's equation by efficient spectral methods~\cite{ISPD16-Lu}. Let $(\omega_j,\omega_k,\omega_l)=(\frac{j\pi}{d_{x}},\frac{k\pi}{d_{y}},\frac{l\pi}{d_{z}})$ denote the frequency indices. The density frequency coefficients $a_{jkl}$ are computed as 
\begin{equation}
  a_{jkl}=\frac{1}{N}\sum_{x,y,z}\rho(x, y, z)\cos(\omega_jx)\cos(\omega_ky)\cos(\omega_lz),
  \label{eq:density-coef}
\end{equation}
where the denominator $N=N_xN_yN_z$ denotes the total number of bins. And according to~\Cref{eq:3d-poisson-equation}, the potential map solution $\phi(x, y, z)$ under constraint $\int_{\Omega}\phi(\bm{v})\,\mathrm{d}\Omega=0$ is given by
\begin{equation}
  \phi(x, y, z)=\sum_{j,k,l}\frac{a_{jkl}}{\omega_j^2+\omega_k^2+\omega_l^2}\cos(\omega_jx)\cos(\omega_ky)\cos(\omega_lz).
  \label{eq:poisson-sol-potential}
\end{equation}
By differentiating~\Cref{eq:poisson-sol-potential}, we have the electric field $\bm{E}(x,y,z)=(E_x,E_y,E_z)$ shown as below
\begin{equation}
  \begin{aligned}
    E_x&=\sum_{j,k,l}\frac{a_{jkl}\omega_j}{\omega_j^2+\omega_k^2+\omega_l^2}\sin(\omega_jx)\cos(\omega_ky)\cos(\omega_lz),\\
    E_y&=\sum_{j,k,l}\frac{a_{jkl}\omega_k}{\omega_j^2+\omega_k^2+\omega_l^2}\cos(\omega_jx)\sin(\omega_ky)\cos(\omega_lz),\\
    E_z&=\sum_{j,k,l}\frac{a_{jkl}\omega_l}{\omega_j^2+\omega_k^2+\omega_l^2}\cos(\omega_jx)\cos(\omega_ky)\sin(\omega_lz).
  \end{aligned}
  \label{eq:possion-sol-force}
\end{equation}
The above spectral equations can be efficiently solved using fast Fourier transform (FFT) with $O(N\log{N})$ complexity.
However, during the forward phase, we need to compute the density map $\rho(x, y, z)$, and during the backward phase, the electric force is $\nabla\hat{U}_i = q_i\bm{E}_i$, which both require the density accumulation over the 3D grid bins.
As a result, the density accumulation becomes the runtime bottleneck.

{
Two types of dummy fillers are inserted into our placement system.
(1) $z$-Fixed Fillers: These fillers are used to manage maximum utilization constraints and maintain fixed $z$-coordinates.
}
Fillers on the same die are equally sized (cuboid) with depth $d = \frac{1}{2}d_{z}$.
We set the total volume of top {$z$-fixed fillers} $\mathrm{vol}_f^+$ and bottom {$z$-fixed fillers} $\mathrm{vol}_f^-$ as 
\begin{equation}
  \begin{aligned}
    \mathrm{vol}_f^+&=\tfrac{1}{2}d_xd_yd_z(1 - u^{+}),\\
    \mathrm{vol}_f^-&=\tfrac{1}{2}d_xd_yd_z(1 - u^{-}),
  \end{aligned}
  \label{eq:filler-vol}
\end{equation}
where $u^{+}$ and $u^{-}$ are the maximum utilization rate for the top die and the bottom die, respectively. 
The top {$z$-fixed fillers} are initialized with $z_i = \frac{3d_z}{4}$, and bottom {$z$-fixed fillers} are initialized with $z_i = \frac{d_z}{4}$.
During the optimization, these fillers' $z$-gradients are set to zero.
Once a die's maximum utilization rate is exceeded, the fillers will push the instances to the other die.
{
(2) Free Fillers: To address the potential white space, we insert free fillers that are initialized at the center of the region $\Omega$ and follow a normal distribution.
Unlike the $z$-fixed fillers, free fillers can move freely in the $z$-direction.
The total volume of free fillers $\mathrm{vol}_{fr}$ is calculated as
\begin{equation}
  \mathrm{vol}_{fr} = \max\left\{d_xd_yd_z - (\mathrm{vol}_f^+ + \mathrm{vol}_f^- + \mathrm{vol}_a), 0 \right\},
  \label{eq:free-filler-vol}
\end{equation}
where $\mathrm{vol}_a$ is the total volume of all the instances.
We randomly initialize all the instances at the center following a normal distribution.
The total instance volume $\mathrm{vol}_a$ is calculated based on this initial position.
}


\begin{figure}[t]
  \centering
  \begin{tikzpicture}[
  net/.style={line width=1pt,densely dashed},
  hbt/.style={rectangle,fill=bole!50,draw=bole}
]
  \draw[net,chazred] (0,0)--(.5,0);
  \node[label=0:{top net $e^{+}$}] at (.5,0) {};
  \draw[net,chazblue] (3,0)--(3.5,0);
  \node[label=0:{bottom net $e^{-}$}] at (3.5,0) {};
  \node[hbt,label=0:{HBT}] (hbt) at (6.5,0) {};
\end{tikzpicture}\\
\begin{subfigure}[]{.49\linewidth}
  \begin{tikzpicture}[
      line width=1pt,
      pin/.style={circle,fill,scale=0.7},
      hbt/.style={rectangle,fill=bole!50,draw=bole},
      btmpin/.style={pin,fill=chazblue!50,draw=chazblue},
      toppin/.style={pin,fill=chazred!50,draw=chazred},
      net/.style={line width=1pt,dashed,opacity=.6},
      seg/.style={line width=1pt,opacity=.6},
      darrow/.style={{Triangle[width=4pt,length=4pt]}-{Triangle[width=4pt,length=4pt]}}
    ]
    \draw[seg] (0,1)--(0,1.5);
    \draw[seg] (2,1)--(2,1.25);
    \draw[seg] (3,0.5)--(3,1.5);
    \draw[seg] (1,-1)--(1,-1.25);
    \draw[seg] (3,-1)--(3,-1.25);
    \draw[seg,darrow] (0,1.25)--(2,1.25);
    \draw[seg,darrow] (0,1.5)--(3,1.5);
    \draw[seg,darrow] (1,-1.25)--(3,-1.25);
    \node[toppin,label=-90:$p_1$] (p1) at (0,1) {};
    \node[btmpin,label=180:$p_2$] (p2) at (1,-1) {};
    \node[toppin,label=90:$p_3$] (p3) at (2,0) {};
    \node[btmpin,label=0:$p_4$] (p4) at (3,0.5) {};
    \node[hbt] (hbt) at (1,0.5) {};
    \begin{pgfonlayer}{background}
      \filldraw[net,draw=chazblue,fill=chazblue!60] (p2) rectangle (p4);
      \filldraw[net,draw=chazred,fill=chazred!60] (p1) rectangle (p3);
    \end{pgfonlayer}
    \node[inner sep=0,label=0:$p_{e^{+}}(\bm{x})$] at (0.5, 1.05) {};
    \node[inner sep=0,label=0:$p_{e^{-}}(\bm{x})$] at (1.5, -1.05) {};
    \node[inner sep=0,label=0:$p_{e}(\bm{x})$] at (1.1, 1.7) {};
  \end{tikzpicture}
  \caption{}
  \label{subfig:hpwl-example-a}
\end{subfigure}
\begin{subfigure}[]{.49\linewidth}
  \begin{tikzpicture}[
    line width=1pt,
    pin/.style={circle,fill,scale=0.7},
    hbt/.style={rectangle,fill=bole!50,draw=bole},
    btmpin/.style={pin,fill=chazblue!50,draw=chazblue},
    toppin/.style={pin,fill=chazred!50,draw=chazred},
    net/.style={line width=1pt,dashed,opacity=.6},
    seg/.style={line width=1pt,opacity=.6},
    darrow/.style={{Triangle[width=4pt,length=4pt]}-{Triangle[width=4pt,length=4pt]}}
  ]
  \draw[seg] (0,1)--(0,1.5);
  \draw[seg] (1,1)--(1,1.25);
  \draw[seg] (3,0.5)--(3,1.5);
  \draw[seg] (1,-1)--(1,-1.25);
  \draw[seg] (3,-1)--(3,-1.25);
  \draw[seg,darrow] (0,1.25)--(1,1.25);
  \draw[seg,darrow] (0,1.5)--(3,1.5);
  \draw[seg,darrow] (1,-1.25)--(3,-1.25);
  \node[toppin,label=-90:$p_1$] (p1) at (0,1) {};
  \node[btmpin,label=180:$p_2$] (p2) at (1,-1) {};
  \node[btmpin,label=90:$p_3$] (p3) at (2,0) {};
  \node[btmpin,label=0:$p_4$] (p4) at (3,0.5) {};
  \node[hbt] (hbt) at (1,0.5) {};
  \begin{pgfonlayer}{background}
    \filldraw[net,draw=chazblue,fill=chazblue!60] (p2) rectangle (p4);
    \filldraw[net,draw=chazred,fill=chazred!60] (p1) rectangle (hbt);
  \end{pgfonlayer}
  \node[inner sep=0,label=0:\footnotesize$p_{e^{+}}(\bm{x})$] at (0., 1.05) {};
  \node[inner sep=0,label=0:$p_{e^{-}}(\bm{x})$] at (1.5, -1.05) {};
  \node[inner sep=0,label=0:$p_{e}(\bm{x})$] at (1.1, 1.7) {};
\end{tikzpicture}
  \caption{}
  \label{subfig:hpwl-example-b}
\end{subfigure}
  \caption{Illustration of die-to-die HPWL wirelength in $x$-axis, the $y$-axis is similar.
    \subref{subfig:hpwl-example-a}~The 3D HPWL is inconsistent with the D2D HPWL. The $x$-axis D2D HPWL is larger than the $x$-axis HPWL of the entire bounding box.
    \subref{subfig:hpwl-example-b}~With the planar locations fixed, changing the pin partition can significantly reduce the $x$-axis D2D HPWL in some cases.
  }
  \label{fig:hpwl-analysis}
\end{figure}
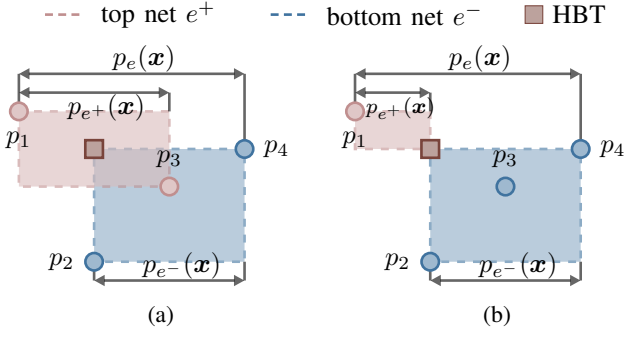
\vskip .5em%

\minisection{\labeltext[3.1]{Bistratal Wirelength Model}{custom:wl-model}}.
According to the objective in~\Cref{eq:high-level-formulation}, the primary optimization goal is to minimize the D2D wirelength in~\Cref{eq:exact-wirelength}, and a small hybrid bonding terminal cost $\beta$ is specified by the design to encourage more usage of HBTs.
The HBT cost can be naturally modeled by $p_{e}(\bm{z})$, reflecting the cut size.
However, the traditional 3D HPWL model in~\Cref{eq:3d-hpwl-def} cannot match the D2D wirelength in~\Cref{eq:exact-wirelength} for the planar wirelength, which contributes most to the objective.

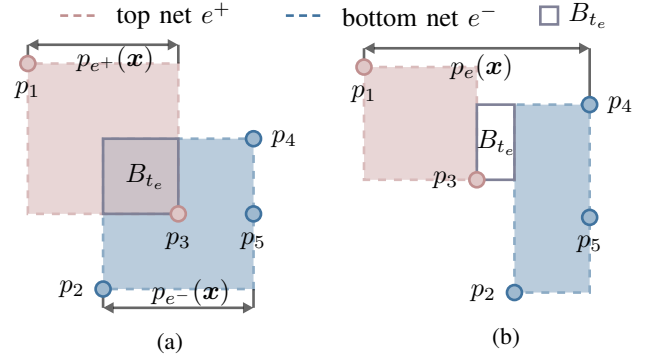
\begin{figure}[t]
  \centering
  \begin{tikzpicture}[
  net/.style={line width=1pt,densely dashed},
  region/.style={rectangle,line width=1pt,draw=chazred!50!chazblue}
]
  \draw[net,chazred] (0,0)--(.5,0);
  \node[label=0:{top net $e^{+}$}] at (.5,0) {};
  \draw[net,chazblue] (3,0)--(3.5,0);
  \node[label=0:{bottom net $e^{-}$}] at (3.5,0) {};
  \node[region,label=0:{$B_{t_e}$}] (region) at (6.5,0) {};
\end{tikzpicture}\\
\begin{subfigure}[]{.49\linewidth}
  \begin{tikzpicture}[
      line width=1pt,
      pin/.style={circle,fill,scale=0.6},
      btmpin/.style={pin,fill=chazblue!50,draw=chazblue},
      toppin/.style={pin,fill=chazred!50,draw=chazred},
      net/.style={line width=1pt,dashed,opacity=.6},
      seg/.style={line width=1pt,opacity=.6},
      darrow/.style={{Triangle[width=4pt,length=4pt]}-{Triangle[width=4pt,length=4pt]}}
    ]
    \draw[seg] (0,2)--(0,2.25);
    \draw[seg] (2,2)--(2,2.25);
    \draw[seg] (1,-1)--(1,-1.25);
    \draw[seg] (3,-1)--(3,-1.25);
    \draw[seg,darrow] (0,2.25)--(2,2.25);
    \draw[seg,darrow] (1,-1.25)--(3,-1.25);
    \draw[draw=chazred!50!chazblue]
    (1,1) rectangle (2,0);
    \node[toppin,label=-90:$p_1$] (p1) at (0,2) {};
    \node[btmpin,label=180:$p_2$] (p2) at (1,-1) {};
    \node[toppin,label=-90:$p_3$] (p3) at (2,0) {};
    \node[btmpin,label=0:$p_4$] (p4) at (3,1) {};
    \node[btmpin,label=-90:$p_5$] (p5) at (3,0) {};
    \begin{pgfonlayer}{background}
      \filldraw[net,draw=chazblue,fill=chazblue!60] (p2) rectangle (p4);
      \filldraw[net,draw=chazred,fill=chazred!60] (p1) rectangle (p3);
    \end{pgfonlayer}
    \node[inner sep=0,label=0:$p_{e^{+}}(\bm{x})$] at (0.5, 2.05) {};
    \node[inner sep=0,label=0:$p_{e^{-}}(\bm{x})$] at (1.5, -1.05) {};
    \node[inner sep=0,label=0:$B_{t_e}$] at (1.15, 0.5) {};
  \end{tikzpicture}
  \caption{}
  \label{subfig:optimal-region-a}
\end{subfigure}
\begin{subfigure}[]{.49\linewidth}
  \begin{tikzpicture}[
    line width=1pt,
    pin/.style={circle,fill,scale=0.6},
    btmpin/.style={pin,fill=chazblue!50,draw=chazblue},
    toppin/.style={pin,fill=chazred!50,draw=chazred},
    net/.style={line width=1pt,dashed,opacity=.6},
    seg/.style={line width=1pt,opacity=.6},
    darrow/.style={{Triangle[width=4pt,length=4pt]}-{Triangle[width=4pt,length=4pt]}}
  ]
    \draw[seg] (0,2)--(0,2.25);
    \draw[seg] (3,1.5)--(3,2.25);
    \draw[seg,darrow] (0,2.25)--(3,2.25);
    \draw[draw=chazred!50!chazblue]
    (1.5,0.5) rectangle (2,1.5);
    \node[toppin,label=-90:$p_1$] (p1) at (0,2) {};
    \node[btmpin,label=180:$p_2$] (p2) at (2,-1) {};
    \node[toppin,label=180:$p_3$] (p3) at (1.5,0.5) {};
    \node[btmpin,label=0:$p_4$] (p4) at (3,1.5) {};
    \node[btmpin,label=-90:$p_5$] (p5) at (3,0) {};
    \begin{pgfonlayer}{background}
      \filldraw[net,draw=chazblue,fill=chazblue!60] (p2) rectangle (p4);
      \filldraw[net,draw=chazred,fill=chazred!60] (p1) rectangle (p3);
    \end{pgfonlayer}
    \node[inner sep=0,label=0:$p_{e}(\bm{x})$] at (1, 2) {};
    \node[inner sep=0,label=0:$B_{t_e}$] at (1.35, 1) {};
  \end{tikzpicture}
  \caption{}
  \label{subfig:optimal-region-b}
\end{subfigure}
  \caption{The optimal region $B_{t_e}$ is the region bounded by the median values of the top net box $B_{e}^{+}$ and bottom net box $B_{e}^{-}$. 
  HBT $t_e$ placed outside $B_{t_e}$ will introduce extra wirelength.
  \subref{subfig:optimal-region-a}~If $B_{e}^{+}$ and $B_{e}^{-}$ overlap in $x$-axis, the minimal $x$-axis D2D HPWL is $p_{e^{+}}(\bm{x})+p_{e^{-}}(\bm{x})$.
  \subref{subfig:optimal-region-b}~If $B_{e}^{+}$ and $B_{e}^{-}$ have no overlap in $x$-axis, the minimal $x$-axis D2D HPWL is $p_{e}(\bm{x})$.
  }
  \label{fig:optimal-region}
\end{figure}

As illustrated in~\Cref{subfig:hpwl-example-a}, the $x$-axis 3D HPWL $p_e(\bm{x})$ is smaller than $x$-axis D2D HPWL $p_{e^+}(\bm{x}) + p_{e^-}(\bm{x})$. 
In fact, the D2D HPWL can be 2$\times$ of the 3D HPWL if the bounding boxes of the top net and the bottom net are the same, and only if the bounding boxes have no overlap as shown in~\Cref{subfig:hpwl-example-b}, they are of the same value.
The error of the 3D HPWL arises from the negligence of the HBTs in D2D placement.
To decide the locations of the HBTs, we first introduce the optimal region for an HBT $t_e$.
Given the bounding box $B_e^{+}=[x_{\min}^{+},x_{\max}^{+}]\times[y_{\min}^{+},y_{\max}^{+}]$ for the top net and $B_e^{-}=[x_{\min}^{-},x_{\max}^{-}]\times[y_{\min}^{-},y_{\max}^{-}]$ for the bottom net, the optimal region $B_{t_e}=[x_{\min}',x_{\max}']\times[y_{\min}',y_{\max}']$ for the HBT $t_e$ is defined as,
\begin{equation}
  \begin{aligned}
    x_{\min}'&=\makebox[\widthof{$\max$}][c]{$\min$}\left\{\max\left\{x_{\min}^{+},x_{\min}^{-}\right\},
    \min\left\{x_{\max}^{+},x_{\max}^{-}\right\}\right\},\\
    x_{\max}'&=\max\left\{\max\left\{x_{\min}^{+},x_{\min}^{-}\right\},
    \min\left\{x_{\max}^{+},x_{\max}^{-}\right\}\right\},
  \end{aligned}
  \label{eq:optimal-region}
\end{equation}
and $y_{\min}',y_{\max}'$ are defined similarly.
With the HBT in its optimal region, the D2D HPWL is minimized as illustrated in~\Cref{fig:optimal-region}.

Based on the above analysis, we can derive the minimal D2D wirelength in $x$-axis, 
\begin{equation}
  W_{e_x}(\bm{x})=\max\left\{p_e(\bm{x}),\ p_{e^{+}}(\bm{x})+p_{e^{-}}(\bm{x})\right\}.
  \label{eq:wl-obj}
\end{equation}
\Cref{eq:wl-obj} demonstrates how to explicitly optimize D2D wirelength in 3D global placement.
If the bounding boxes $B_e^{+}$ and $B_e^{-}$ overlap, we optimize the HPWL of each partial net as shown in~\Cref{subfig:optimal-region-a}. 
Otherwise, we optimize the entire bounding box at the non-overlapping direction as shown in~\Cref{subfig:optimal-region-b}. 
Combining $y$-axis wirelength which is calculated similarly, the bistratal wirelength~\cite{TCAD23-Liao} is defined by $W_{e,\text{Bi}}(\bm{x},\bm{y},\bm{z})=W_{e_x}(\bm{x})+W_{e_y}(\bm{y})$. 
It is worth mentioning that $W_{e,\text{Bi}}(\cdot)$ is also a function of $\bm{z}$ as $z$-coordinates determine partial nets $e^+, e^-$ directly. 
Meanwhile, we also dynamically update the pin offset values in the same approach as~\Cref{eq:std-size-tech} to model the heterogeneous technology nodes.
Combining the HBT cost, our wirelength model for the 3D global placement is
\begin{equation}
  W_{e}(\bm{x}, \bm{y}, \bm{z})=W_{e,\text{Bi}}(\bm{x},\bm{y},\bm{z}) + \alpha p_{e}(\bm{z}).
  \label{eq:wl-model}
\end{equation}

Applying the weighted-average model in~\Cref{eq:wa-model-approx} to $p_e(\cdot)$, we can perform gradient-based optimization on the smoothed objective $\hat{W}_e$.
However, the weighted-average model only minimizes $p_{e}(\bm{z})$ to reduce the cut size, incapable of optimizing the partition for wirelength. 
As shown in~\Cref{subfig:hpwl-example-b}, the D2D wirelength can be greatly reduced with a better distribution of $\bm{z}$.
The smoothed bistratal wirelength $\hat{W}_{e,\text{Bi}}$ is discontinuous with respect to $\bm{z}$, therefore, the gradient $\nabla_{\bm{z}}W_{e,\text{Bi}}$ does not exist.
To approximate the impact of $\bm{z}$ on wirelength, we leverage finite difference approximation~\cite{TCAD23-Liao,milne2000calculus} to perform numerical optimization on $\bm{z}$ with gradient defined by
{
\begin{multline}
  (\nabla_{\bm{z}}\hat{W}_{e,\text{Bi}})_i=\tfrac{4}{d_z}\Big(W_{e,\text{Bi}}(\bm{x}, \bm{y},\tilde{\bm{z}}_i+\tfrac{3d_z}{4}\bm{e}_i)\\
  -W_{e,\text{Bi}}(\bm{x}, \bm{y}, \tilde{\bm{z}}_i+\tfrac{d_z}{4}\bm{e}_i)\Big),
  \label{eq:fda-grad}
\end{multline}
}%
where $\bm{x}$ and $\bm{y}$ are fixed for wirelength evaluation, and $\tilde{\bm{z}}_i=\bm{z}\odot(\bm{1}-\bm{e}_i)$ and $\bm{e}_i\in\mathbb{R}^{|V|}$ is the unit vector with the $i$-th entry being 1 and others being 0.
For each instance, we perturb $z_i$ with $\Delta z = \frac{d_z}{4}$ to alter its partition and evaluate the bistratal wirelength change $\Delta W_{e, \text{Bi}}$.
The difference quotient is adopted as the gradient, which provides a local view of wirelength benefits for updating $z_{i}$.

{
To better demonstrate the derivation of~\Cref{eq:fda-grad}, we analyze $z_i$ within interval $[\frac{d_z}{4},\frac{3d_z}{4}]$ for instance $v_i\in V$ and fix all other variables. 
Then, $W_{e,\text{Bi}}(\bm{x},\bm{y},\bm{z})$ simplifies to $W_{e,\text{Bi}}(z_i)$ which is a step function that only takes two possible values $W_{e,\text{Bi}}(\frac{d_z}{4})$ and $W_{e,\text{Bi}}(\frac{3d_z}{4})$, corresponding to the wirelength when instance $v_i$ is on the bottom and top dies, respectively. 
Specifically, we always have $W_{e,\text{Bi}}(z_i)=W_{e,\text{Bi}}(\frac{d_z}{4})$ if $z_i<\frac{d_z}{2}$ and $W_{e,\text{Bi}}(z_i)=W_{e,\text{Bi}}(\frac{3d_z}{4})$ otherwise.
In fact, the finite difference approximation computes the partial derivative $(\nabla_{\bm{z}}\hat{W}_{e,\text{Bi}})_i$ as follows for $z_i\in[\frac{1}{4}d_z,\frac{1}{2}d_z)$
\begin{equation}
  \begin{aligned}
    \underset{\frac{1}{4}d_z}{\Delta}W_{e,\text{Bi}}(z_i)&=\frac{W_{e,\text{Bi}}(z_i+\tfrac{1}{4}d_z)-W_{e,\text{Bi}}(z_i)}{\tfrac{1}{4}d_z}\\
    &=\tfrac{4}{d_z}\left(W_{e,\text{Bi}}(\tfrac{3d_z}{4})-W_{e,\text{Bi}}(\tfrac{d_z}{4})\right)
  \end{aligned}
  \label{eq:fda-deriv-positive}
\end{equation}
Similarly, if $z_i\in[\frac{1}{2}d_z,\frac{3}{4}d_z]$, it computes
\begin{equation}
  \begin{aligned}
    \underset{-\frac{1}{4}d_z}{\Delta}W_{e,\text{Bi}}(z_i)&=\frac{W_{e,\text{Bi}}(z_i-\tfrac{1}{4}d_z)-W_{e,\text{Bi}}(z_i)}{-\tfrac{1}{4}d_z}\\
    &=\tfrac{4}{d_z}\left(W_{e,\text{Bi}}(\tfrac{3d_z}{4})-W_{e,\text{Bi}}(\tfrac{d_z}{4})\right).
  \end{aligned}
  \label{eq:fda-deriv-negative}
\end{equation}
Combining~\Cref{eq:fda-deriv-positive} and~\Cref{eq:fda-deriv-negative}, we obtain~\Cref{eq:fda-grad} as a conclusion.
}

\minisection{\labeltext[3.1]{3D Mixed-Size Preconditioning}{custom:precond}}.
Preconditioning is a critical component of numerical optimization which reduces the condition number and stabilizes the optimization process. 
The large topological and physical difference between macros and standard cells makes the preconditioner indispensable in nonlinear placement optimization.

\Cref{eq:fda-grad} provides the optimization direction for $\bm{z}$. 
However, $\nabla_{\bm{z}}\hat{W}_{e,\text{Bi}}$ is not on the same scale as planar gradients $\nabla_{\bm{x}}\hat{W}_{e,\text{Bi}}$ and $\nabla_{\bm{y}}\hat{W}_{e,\text{Bi}}$, leading to suboptimal results. 
Hence, we normalize~\Cref{eq:fda-grad} before applying gradient descent,
\begin{equation}
  \bm{g}=\frac{\|\nabla_{\bm{x}}\hat{W}_{e,\text{Bi}}\|_1+\|\nabla_{\bm{y}}\hat{W}_{e,\text{Bi}}\|_1}{2\|\nabla_{\bm{z}}\hat{W}_{e,\text{Bi}}\|_1}\nabla_{\bm{z}}\hat{W}_{e,\text{Bi}},
  \label{eq:fda-normalize}
\end{equation}
and we use $(\nabla_{\bm{x}}\hat{W}_{e,\text{Bi}},\nabla_{\bm{y}}\hat{W}_{e,\text{Bi}},\bm{g} + \alpha \nabla_{\bm{z}} \hat{p}_e(\bm{z}))$ as the gradient of our wirelength objective in 3D mixed-size preconditioning, {which ensures the continuity of the optimization process.}

{In 2D placement, ePlace~\cite{TODAES2015-Lu-ePlace, PLACE-TCAD2015-Lu} adopts the Jacobi preconditioner}, which only selects diagonal entries of the Hessian matrix, to perform preconditioning on gradients.
Let $f(\bm{v})$ be the objective function in~\Cref{eq:analytical-placement}. 
Considering $x$ direction, the $i$-th diagonal entry of Hessian matrix $\bm{H}_f=\nabla_{\bm{x}}^2f$ is given by
\begin{equation}
  (\bm{H}_f)_{ii}=\sum_e\frac{\partial^2\hat{W}_e}{\partial x_i^2}+\lambda\frac{\partial^2\hat{U}}{\partial x_i^2}.
  \label{eq:general-precond}
\end{equation}
{ePlace~\cite{TODAES2015-Lu-ePlace, PLACE-TCAD2015-Lu} approximates~\Cref{eq:general-precond} with $\sum_e\frac{\partial^2\hat{W}_e}{\partial x_i^2} \approx |E_i|$, $ \frac{\partial^2\hat{U}}{\partial x_i^2} \approx q_i$, and $(\bm{H}_f)_{ii} \approx |E_i| + \lambda q_i$, for both standard cells and macros}, where $|E_i|$ is the set cardinality of all nets incident to instance $v_i\in V$ and $q_i$ stands for the electric quantity of $v_i$. 
Specifically, $|E_i|$ is the number of pins on $v_i$, and $q_i$ is the corresponding instance area or volume. 
To better adapt to the third dimension, {ePlace3D}~\cite{ISPD16-Lu} removes the first item and only uses {$(\bm{H}_f)_{ii} \approx \lambda{q_i}$} as the preconditioner {for both standard cells and macros}.

However, we have $\lambda q_i\ll1$ at the early global placement stage when the density weight $\lambda$ is small, resulting in a stability issue and subsequent divergence.
The wirelength gradients of macros are significantly larger than those of standard cells, as the macros have a larger number of pins.
The large movement of macros frequently perturbs the optimization direction.
Therefore, we propose the 3D mixed-size preconditioner as follows,
\begin{equation}
  (\bm{H}_f)_{ii}\approx
  \left\{
    \begin{array}{ll}
      \max\left\{1,|E_i|+\lambda q_i\right\}, &\text{if }v_i\in V_M,\\
      \max\left\{1,\lambda q_i\right\}, &\text{otherwise}.\\
    \end{array}
  \right.
  \label{eq:improved-precond}
\end{equation}
Through the mixed-size preconditioning in~\Cref{eq:improved-precond}, the macros are allowed to move at the pace of standard cells at the early global placement stage, preventing the optimization from divergence.
With density weight $\lambda$ increasing, the spreading standard cells provide enough physical information to drive the macros to the proper die.

\begin{table}[tb!]
    \centering
    \caption{Notations for the MILP formulation of macro rotation assignment.}
    \resizebox{\linewidth}{!}{
      \begin{tabular}{c|l}
        \toprule
        \textbf{Notations} & \multicolumn{1}{c}{\textbf{Descriptions}}\\
        \midrule
        $S_{j}$ & a set of standard cell instances connected by $e_j$\\
        $M_{j}$ & a set of macro instances connected by $e_j$\\
        $(x_i, y_i)$ &  center location of instance $v_i$\\
        $(o_{ij}^{x}, o_{ij}^{y})$ & pin offsets on $v_i$ connected by $e_j$ with respect to the center of $v_i$\\
        $(r_i, r'_i)$ &  binary variables to encode the rotation of instance $v_i$\\
        \bottomrule
      \end{tabular}
    }
    \label{tab:macro-rotation}
\end{table}
\subsection{MILP Macro Rotation Assignment}
\label{subsec:macro-rotation}
The initial 3D placement solution provides valuable information about the locations and partition of the macros and standard cells.
Based on the physical information, we propose a mixed-integer linear programming (MILP) formulation to assign macro rotations to minimize wirelength.

We only need to consider the net set $E_M$ connecting the macros $V_M$ to find the optimal rotation assignment.
The notations used in the formulation are summarized in~\Cref{tab:macro-rotation}.
Consider arbitrary net $e_j\in E_M$ connecting a set of instances including a set of standard cells $S_j$ and a set of macros $M_j$.
\iffalse
We use $(o_{ij}^{x},o_{ij}^{y})$ to denote the pin offset values on instance $v_i$ connected by $e_j$, and determine the pin coordinate $(x_{ij},y_{ij}):=(x_i+o_{ij}^{x}, y_i+o_{ij}^{y})$ accordingly.
\else
We use $(o_{ij}^{x},o_{ij}^{y})$ to denote the pin offset values on instance $v_i$ connected by $e_j$.
\fi
For standard cell $v_k \in S_j$, which cannot be rotated, the coordinates of the pin on $v_k$ connecting to $e_j$ are given by $(x_{kj},y_{kj}):=(x_k+o_{kj}^{x}, y_k+o_{kj}^{y})$.
For the pin location of rotatable macro $v_i \in M_j$ connecting to $e_j$, we use two binary variables to represent its coordinates $(x_{ij},y_{ij})$:
\begin{equation}
  \begin{aligned}
    \iffalse
    \begin{bmatrix}\tilde{x}_{ij}\\\tilde{y}_{ij}\end{bmatrix}&=
    \begin{bmatrix}x_i\\y_i\end{bmatrix}+
    \begin{bmatrix}1-r_i-r_i'&r_i-r_i'\\r_i'-r_i&1-r_i-r_i'\end{bmatrix}
    \begin{bmatrix}o_{ij}^x\\o_{ij}^y\end{bmatrix}\\
    &=\begin{bmatrix}x_{ij}\\y_{ij}\end{bmatrix}+
    \begin{bmatrix*}[r]-o_{ij}^x+o_{ij}^y&-o_{ij}^x-o_{ij}^y\\-o_{ij}^x-o_{ij}^y&o_{ij}^x-o_{ij}^y\end{bmatrix*}
    \begin{bmatrix}r_i\\r_i'\end{bmatrix}.
    \else
    x_{ij} &= x_i + (1 - r_i - r'_i)o_{ij}^{x} + (r_i - r'_i)o_{ij}^{y}\ , \\
    y_{ij} &= y_i + (r'_i - r_i)o_{ij}^{x} + (1 - r_i - r'_i)o_{ij}^{y}\ .
    \fi
  \end{aligned}
  \label{eq:macro-pin}
\end{equation}
The binary variables $(r_i, r'_i)$ with values $(0,0), (0,1), (1,1),$ and $(1,0)$ indicate that macro rotates counterclockwise by 0$^{\circ}$, 90$^{\circ}$, 180$^{\circ}$, and 270$^{\circ}$, respectively.

Since the rotation assignment is performed after the initial 3D placement, all the instances are distributed to the corresponding dies according to $z$-coordinates, and HBTs for the crossing-die nets are inserted at the center point of the optimal region in~\Cref{eq:optimal-region}.
The problem is reduced to the 2D scenario.
Our objective is to minimize the total D2D wirelength of net set $E_M$, leading to the following MILP formulation,
\begin{equation}
  \begin{array}{cll}
    \displaystyle\min &\multicolumn{2}{l}{\displaystyle\sum_{e_j\in E_M}(R_j^x-L_j^x+R_j^y-L_j^y)}\\[3\jot]
    \mathrm{s.t.}&
    \iffalse
    \begin{bmatrix}L_j^x\\L_j^y\end{bmatrix}\leq
    \begin{bmatrix}x_{kj}\\y_{kj}\end{bmatrix}\leq
    \begin{bmatrix}R_j^x\\R_j^y\end{bmatrix},&\forall v_k \in S_j\\[3\jot]
    &\begin{bmatrix}L_j^x\\L_j^y\end{bmatrix}\leq
    \begin{bmatrix}x_{ij}\\y_{ij}\end{bmatrix}+
    A_{ij}\cdot\begin{bmatrix}r_i\\r_i'\end{bmatrix}\leq
    \begin{bmatrix}R_j^x\\R_j^y\end{bmatrix},&\forall v_i \in M_j
    \else
    L_j^x \leq x_k + o_{kj}^{x} \leq R_j^x, &\forall v_k \in S_j\\
    & x_i + (1 - r_i - r'_i)o_{ij}^{x} + (r_i - r'_i)o_{ij}^{y} \leq R_j^x, &\forall v_i \in M_j\\
    & x_i + (1 - r_i - r'_i)o_{ij}^{x} + (r_i - r'_i)o_{ij}^{y} \geq L_j^x, &\forall v_i \in M_j\\
    & L_j^y \leq y_k + o_{kj}^{y} \leq R_j^y, &\forall v_k \in S_j\\
    & y_i + (r'_i - r_i)o_{ij}^{x} + (1 - r_i - r'_i)o_{ij}^{y}\leq R_j^y, &\forall v_i \in M_j \\
    & y_i + (r'_i - r_i)o_{ij}^{x} + (1 - r_i - r'_i)o_{ij}^{y} \geq L_j^y, &\forall v_i \in M_j
    \fi
  \end{array}
  \label{eq:MILP}
\end{equation}
$R_j^x\ (R_j^y)$ and $L_j^x\ (L_j^y)$ represent the $x\ (y)$ bounding box boundary to optimize.
Note that we consider the HBT locations by treating the HBT as standard cell at this stage. 
There are $O(|V_M|)$ binary variables and $O(|E_M|)$ linear constraints, which are relatively small.
We can solve it optimally by invoking an MILP solver with negligible runtime overhead.
{
Additionally, our MILP formulation is sufficiently flexible to accommodate the orientation constraints of the latest technology node by disabling the corresponding orientation variables.
The potential instance overlap resulting from macro rotation will be resolved during the global placement at a later stage.
}

\subsection{Multi-Die 2D Global Placement}
\label{subsec:2d-gp}

With the optimized macro rotations, our framework performs global placement again to improve the placement quality.
Although the 3D mixed-size global placement can explore the entire solution space, it has difficulty in finding a good macro placement for the design with excessively large macros.
We propose a multi-die 2D global placement formulation removing $z$-dimension to resolve the issue.
{The instance partition is determined by the initial 3D placement solution.}

We model the top die, bottom die, and bonding terminal layer as independent 2D electrostatic fields~\cite{TODAES2015-Lu-ePlace} so that the partitioned macros can spread more easily without the influence of the macro density obstacle on the other die.
The objective of multi-die 2D global placement is to minimize the D2D wirelength while the instances and HBTs on the three layers have minimal overlap, shown as follows
\begin{equation}
    \min_{\bm{x}, \bm{y}}\sum_{e\in E} \hat{W}_e(\bm{x}, \bm{y})+\langle\bm{\lambda},\hat{\bm{U}}\rangle,
    \label{eq:2d-place}
\end{equation}
where $\bm{\lambda} = (\lambda^{+}, \lambda^{-}, \lambda')$ is the vector of the density weights and $\hat{\bm{U}} = (\hat{U}^{+}, \hat{U}^{-}, \hat{U}')$ is the vector of the density penalty for the top die, bottom die, and HBT layer, respectively.
{
We insert dummy fillers into the placement system.
The total area of fillers is determined by subtracting the total instance area from the die area. 
}
The independent 2D density models give more flexibility for the macros compared to the 3D density model, and the HBTs, connecting the top partial nets and the bottom partial nets, guide the connected instances to align in an F2F manner during the placement.

\subsection{Legalization}
\label{subsec:legal}
Die-by-die legalization is performed for macros, standard cells, and HBTs to remove the overlap.
We utilize the transitive closure graph (TCG)~\cite{lin2001tcg, chen2008constraint} to represent the relation between macros, 
and the dual problem of TCG-based macro legalization is associated with the min-cost flow problem~\cite{lin2017mrdp}, which can be solved efficiently by the network simplex algorithm.
We legalize the standard cells die-by-die with Tetris~\cite{hill2002method} and Abacus~\cite{spindler2008abacus} algorithms.
{
The HBTs share the same square size $w'\times w'$ and require a minimum spacing $s'$ between each other.
Hence, we pad the HBT to a square with size $w'+s'$ and legalize them as ordinary standard cells with row height $w'+s'$.
The actual position of the HBTs can be derived from the padded HBTs.
}

\subsection{Detailed Placement}
\label{subsec:dp}
We adopt ABCDPlace~\cite{lin2020abcdplace} as our detailed placement engine, including strategies of global swap~\cite{popovych2014density}, independent set matching~\cite{chen2008ntuplace3}, and local reordering~\cite{pan2005efficient}.
The instances and HBTs on each layer are refined sequentially.
After one iteration of detailed placement, the optimal regions of HBTs may be changed. 
Hence, we can map the HBT to the center point of the updated optimal region, followed by a new iteration of HBT legalization and detailed placement.
The wirelength improvement is negligible for more iterations of the process.
Therefore, we only perform one additional iteration of the detailed placement.

\section{Density and Wirelength Algorithms}
\label{sec:agl}

\subsection{Adaptive 3D Density Accumulation}
\label{subsec:density-accu}
The density accumulation is computation-intensive, becoming the runtime bottleneck in 3D global placement.
Density accumulation includes two phases: the forward phase to compute the density map $\rho$ from instances, and the backward phase to accumulate the weights from the electric field maps $\bm{E}$ to instances.
Two phases share the same primitive operation to compute the overlapping region of instances and bins.
The computation workload can be very imbalanced for standard cells and macros.
Therefore, adaptive algorithms are desired for mixed-size designs in 3D scenarios.

For an instance $v_i$ with size $w_i\times h_i\times\frac{d_z}{2}$, the corresponding cuboid region is $D_{v_i}=[x_i-\frac{w_i}{2},x_i+\frac{w_i}{2}]\times[y_i-\frac{h_i}{2},y_i+\frac{h_i}{2}]\times[z_i-\frac{d_z}{4},z_i+\frac{d_z}{4}]$.
The density map $\rho$ has a size of $|B|=N_x\times N_y\times N_z$. 
For each bin $b\in B$ as a cuboid with size $w_b\times h_b\times d_b$, the density is calculated as,
\begin{equation}
  \rho_b=\sum_{v_i\in V}\omega_{v_i}\frac{\mathrm{vol}(D_{v_i}\cap b)}{\mathrm{vol}(b)},
  \label{eq:density-map-accum}
\end{equation}
where $\omega_{v_i}$ is the weight of instance $v_i$ and $\mathrm{vol}(\cdot)$ is the volume of the cuboid region.
{
  We implement the local smoothness technique as described in~\cite{TODAES2015-Lu-ePlace}, and $\omega_{v_i}$ is determined by the relative sizes of the instance and the bin, $\omega_{v_i} = \min\left\{1,\frac{w_i}{\sqrt{2}w_b}\right\} \times \min\left\{1, \frac{h_i}{\sqrt{2}h_b}\right\}$.
  Given that all instances have the same depth, we do not apply local smoothness to the depth dimension.
} 
{The approach used in prior work~\cite{TCAD23-Liao, DAC19-Lin} for calculating}~\Cref{eq:density-map-accum} is to allocate one thread for each instance and sequentially update all the overlapped bins within that thread.
However, large macros in 3D scenarios may cover many bins, causing severe load balancing issues.

A natural idea for solving the problem is to exploit different levels of parallelism for standard cells and macros.
{Instance parallelism~\cite{TCAD23-Liao}} for standard cells is abundant, and the workload for each thread is light and balanced.
In contrast to the standard cells, the number of macros is small, but the number of bins to traverse for density calculation is much larger.
The key to achieving efficient macro density accumulation is to effectively exploit the bin parallelism.

{
Guo \textit{et al.}~\cite{DAC2021-Guo-Prefix} applied bin parallelism using the prefix sum algorithm in 2D scenarios. 
Specifically, they decomposed each macro into 4 bottom-right instances, with each instance further divided into 4 sub-instances. 
These sub-instances can be processed as increments on bottom-right submatrices or on individual grids.
However, their decomposition strategy, tailored for the 2D density model, is challenging to extend to 3D scenarios due to its reliance on manual design. In contrast, we propose a general formulation for 3D density accumulation with a theoretical guarantee of correctness.
}

\begin{figure}[tb!]
  \input{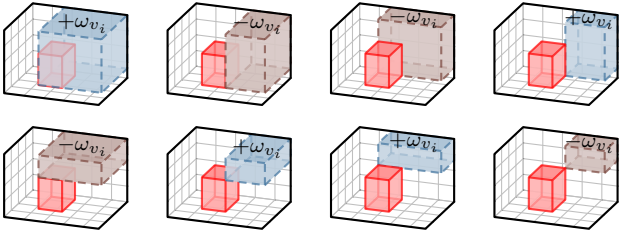}
  \caption{
    The 3D density map calculation for macro $v_i$ colored in \textcolor{red}{red} is decomposed to the weighted sum of 8 corner maps, performed by 3D prefix sum.
    \textcolor{deepblue}{Blue} sub-maps indicate addition region, and \textcolor{bole}{brown} sub-maps indicate subtraction region.
  }
  \label{fig:prefix-sum-cubes}
\end{figure}

The \emph{3D prefix sum} operator is a function $\varphi:\mathbb{R}^{N_x\times N_y\times N_z}\rightarrow\mathbb{R}^{N_x\times N_y\times N_z}$ such that
\begin{equation}
\varphi(\mathcal{A})_{ijk}=\sum_{i'=1}^{i}\sum_{j'=1}^{j}\sum_{k'=1}^{k}\mathcal{A}_{i'j'k'}
\label{eq:def-prefix-sum-3d}
\end{equation}
holds for any 3D map $\mathcal{A}\in\mathbb{R}^{N_x\times N_y\times N_z}$ and valid index tuple $(i,j,k)$.
The prefix sum can propagate a single value in $\mathcal{A}$ to the region with larger indices in time complexity $O(N_xN_yN_z)$.
Based on this idea, we can efficiently compute the density map for macros by only considering the 8 corners, as illustrated in~\Cref{fig:prefix-sum-cubes}.

We first gather the density values at the macro corners. 
Consider a corner $(x,y,z)$ and its normalized coordinates $(\hat{x},\hat{y},\hat{z}):=(\frac{x}{w_b},\frac{y}{h_b},\frac{z}{d_b})$.
Note that a corner may not align precisely with the 3D grid of bins.
And we introduce the function $g(a)=\max\{1-|a|,0\}$ for the partial density introduced by the non-integer coordinates.
Let the 3D map $\mathcal{A}^{(x,y,z)}$ be induced according to the following mechanism,
\begin{equation}
  \mathcal{A}_{ijk}^{(x,y,z)}=
  g\left(i-1-\hat{x}\right)g\left(j-1-\hat{y}\right)g\left(k-1-\hat{z}\right).
  \label{eq:3d-density-accum-corner}
\end{equation}
$\mathcal{A}^{(x,y,z)}$ is sparse with at most 8 non-zero entries adjacent to the bin index {$(\lceil{\hat{x}}\rceil,\lceil{\hat{y}}\rceil,\lceil{\hat{z}}\rceil)$}.
Then, we have the following theorem.

\begin{mytheorem}
  \label{thm:basis-prefix-sum-3d}
  For macro $v_i$ with size $w_i\times h_i\times\frac{d_z}{2}$, center coordinate $(x_i,y_i,z_i)$, and corresponding cuboid region $D_{v_i}$, consider 3D map
  \begin{equation}
    \mathcal{A}_{v_i}=\sum_{\sigma_x,\sigma_y,\sigma_z\in\{-1,1\}}
    -\sigma_x\sigma_y\sigma_z\mathcal{A}^{(x_i+\sigma_x\frac{w_i}{2},y_i+\sigma_y\frac{h_i}{2},z_i+\sigma_z\frac{d_z}{4})}\ .
    \label{eq:3d-density-accum-cuboid}
  \end{equation}
  Then, its prefix sum satisfies $\varphi(\mathcal{A}_{v_i})_b=\frac{\mathrm{vol}(D_{v_i}\cap b)}{\mathrm{vol}(b)}$ for each bin $b\in B$.
\end{mytheorem}

\begin{figure}[tb!]
  \centering
  \begin{subfigure}[b]{.49\linewidth}
    \centering\tiny
    \begin{tikzpicture}[scale=.95,
        font=\scriptsize,
        line width=1pt,
        seg/.style={line width=1pt,opacity=.6},
        pin/.style={circle,fill=bole!50,draw=bole,scale=0.7},
        darrow/.style={{Triangle[width=4pt,length=4pt]}-{Triangle[width=4pt,length=4pt]}}
        ]
        \foreach\x in{1,...,9}{\draw[gray!30] (.5*\x,0.5)--(.5*\x,4.5);}
        \foreach\y in{2,...,9}{\draw[gray!30] (0,.5*\y)--(4.5,.5*\y);}
        \draw (0,0.5) rectangle (4.5, 4.5);
        \filldraw[opacity=.6,fill=myred!60,draw=myred!80!black]
        (0.65,1.6) rectangle (3.65,3.85);

        \draw[seg] (0.65,3.85)--(0.65,4.15);
        \draw[seg] (3.65,3.85)--(3.65,4.15);
        \draw[seg,darrow] (0.65,4.15)--(3.65,4.15);
        \node[inner sep=0] at (2.15, 4.28) {\textbf{6}};

        \draw[seg] (0.65,1.6)--(0.35,1.6);
        \draw[seg] (0.65,3.85)--(0.35,3.85);
        \draw[seg,darrow] (0.35,1.6)--(0.35,3.85);
        \node[inner sep=0, rotate=270] at (0.22, 2.725) {\textbf{4.5}};

        \node[pin, label=270:\textbf{(1.3, 2.2)}] (p1) at (0.65, 1.6) {};

        \node at (0.75, 1.75) {0.56};
        \node at (3.75, 1.75) {0.24};
        \node at (3.75, 3.75) {0.21};
        \node at (0.75, 3.75) {0.49};
        \foreach\x in{3,...,7}{
            \node at ($(.5*\x,1.75)-(0.25,0)$) {0.8};
            \foreach\y in {5,...,7}{
              \node at ($(.5*\x,.5*\y)-(0.25,0.25)$) {1};
            }
            \node at ($(.5*\x,3.75)-(0.25,0)$) {0.7};
        }
        \foreach\y in {5,...,7}{
            \node at ($(0.75,.5*\y)-(0,0.25)$) {0.7};
            \node at ($(3.75,.5*\y)-(0,0.25)$) {0.3};
        }
    \end{tikzpicture}
    \caption{}
    \label{subfig:density-map}
\end{subfigure}\hspace{.15em}
\begin{subfigure}[b]{.49\linewidth}
    \centering\tiny
    \begin{tikzpicture}[scale=.95,font=\scriptsize,line width=1pt]
        \foreach\x in{1,...,9}{\draw[gray!30] (.5*\x,0.5)--(.5*\x,4.5);}
        \foreach\y in{2,...,9}{\draw[gray!30] (0,.5*\y)--(4.5,.5*\y);}
        \draw (0,0.5) rectangle (4.5, 4.5);
        \filldraw[opacity=.6,fill=myred!60,draw=myred!80!black]
        (0.65,1.6) rectangle (3.65,3.85);
        \filldraw[opacity=.6,fill=bole!50,draw=bole]
        (0.5,1.5) rectangle (1,2);

        \node at (0.75,1.75) {0.56}; \node at (1.25,1.75) {0.24};
        \node at (0.75,2.25) {0.14}; \node at (1.25,2.25) {0.06};
        \node at (0.7,3.75) {-0.21}; \node at (1.25,3.75) {-0.09};
        \node at (0.7,4.25) {-0.49}; \node at (1.25,4.25) {-0.21};
        \node at (3.7,1.75) {-0.56}; \node at (4.225,1.75) {-0.24};
        \node at (3.7,2.25) {-0.14}; \node at (4.225,2.25) {-0.06};
        \node at (3.75,3.75) {0.21}; \node at (4.25,3.75) {0.09};
        \node at (3.75,4.25) {0.49}; \node at (4.25,4.25) {0.21};

        \node[inner sep=0] at (2.25, 1.2) {${\mathcal{A}_v}_{2,3} = (-1)\times(-1)\times \mathcal{A}_{2,3}^{(1.3,2.2)}$};
        \node[inner sep=0] at (2.28, .75) {$ = g(-0.3) \times g(-0.2)$};
    \end{tikzpicture}
    \caption{}
    \label{subfig:prefix-map}
\end{subfigure}
  \label{fig:prefix-exp}
  \caption{A 2D density accumulation example using our prefix sum approach. \subref{subfig:density-map} The resulting density map $\rho$. Entries without a number represent zero.\subref{subfig:prefix-map} The 2D map $\mathcal{A}_v$ calculated using~\Cref{eq:3d-density-accum-corner} and~\Cref{eq:3d-density-accum-cuboid}. Apparently, the prefix sum of $\mathcal{A}_v$ equals the density map $\rho$. 3D density accumulation follows a similar rationale.}
  \label{fig:prefix-example}
\end{figure}
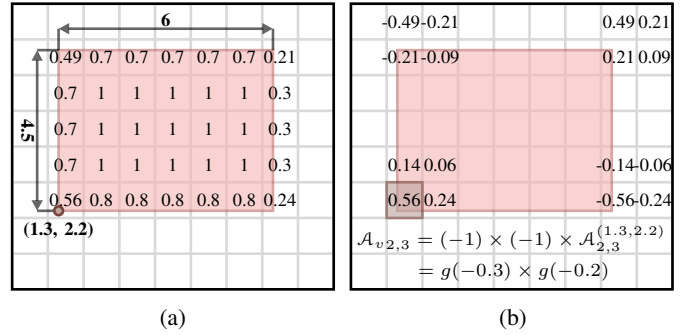

\Cref{thm:basis-prefix-sum-3d} demonstrates the way to simplify the 3D density accumulation for macros. 
With $\sigma_x,\sigma_y,\sigma_z$ as binary variables, and the maps being sparse with 8 real numbers, 
the summation in~\Cref{eq:3d-density-accum-cuboid} can be finished in constant time. 
The calculation of density map for macros is extremely fast by adding $\mathcal{A}_{v_i}$ for all macros $V_M$ followed by a single time of 3D prefix sum.
Therefore, the prefix sum density accumulation runs in $O(N_xN_yN_z+|V_M|)$, which is linear in both the number of bins and macros.
{
\Cref{fig:prefix-example} illustrates a 2D density accumulation example using our prefix sum approach. 
The prefix sum of $\mathcal{A}_{v_i}$, calculated according to~\Cref{eq:3d-density-accum-corner} and~\Cref{eq:3d-density-accum-cuboid}, equals the density map $\rho$. 
And 3D density accumulation follows a similar rationale.
}

In the backward phase, each instance receives the electric force from the overlapped bins in 3 directions, which requires performing prefix sum on electric field maps $\bm{E}$.
The procedure is similar to the forward density accumulation.
The 3D prefix sum is a one-time cost, and the electric force for each macro can be induced at the 8 corners with constant-time summation operations.
Hence, the time complexity for the backward phase is also $O(N_xN_yN_z+|V_M|)$.

Our adaptive method utilizes the instance parallelism for the standard cells and bin parallelism for the macros, which reduces the runtime of 3D global placement from 400s to 157s on \texttt{case4} of the ICCAD 2023 contest benchmarks~\cite{ICCAD23-Hu} {compared to the approach~\cite{TCAD23-Liao}}.


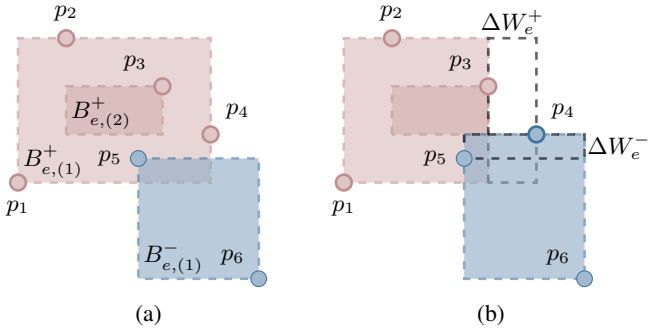
\begin{figure}
  \centering\small
  \begin{subfigure}[b]{.45\linewidth}
  \begin{tikzpicture}[
      line width=1pt,scale=.64,
      pin/.style={circle,fill,scale=0.7},
      hbt/.style={rectangle,fill=bole!50,draw=bole},
      btmpin/.style={pin,fill=chazblue!50,draw=chazblue},
      toppin/.style={pin,fill=chazred!50,draw=chazred},
      net/.style={line width=1pt,dashed,opacity=.6},
      seg/.style={line width=1pt,opacity=.6},
      darrow/.style={{Triangle[width=4pt,length=4pt]}-{Triangle[width=4pt,length=4pt]}}
    ]
    \node[toppin,label=-90:$p_1$] (p1) at (0,0) {};
    \node[toppin,label=90:$p_2$] (p2) at (1,3) {};
    \node[toppin,label=135:$p_3$] (p3) at (3,2) {};
    \node[toppin,label=45:$p_4$] (p4) at (4,1) {};
    \path let\p2=(p2),\p4=(p4) in coordinate (tr1) at (\x4,\y2);
    \path let\p2=(p2),\p4=(p4) in coordinate (bl2) at (\x2,\y4);
    \begin{pgfonlayer}{background}
    \filldraw[net,draw=chazred,fill=chazred!60] (p1) rectangle (tr1);
    \filldraw[net,draw=chazred,fill=chazred!70] (bl2) rectangle (p3);
    \end{pgfonlayer}
    \node[btmpin,label=180:$p_5$] (p5) at (2.5,0.5) {};
    \node[btmpin,label=135:$p_6$] (p6) at (5,-2) {};
    \path let\p5=(p5),\p6=(p6) in coordinate (blb1) at (\x5,\y6);
    \begin{pgfonlayer}{background}
    \filldraw[net,draw=chazblue,fill=chazblue!60] (p5) rectangle (p6);
    \end{pgfonlayer}
    \node[inner sep=0,label={[label distance=-1mm]45:$B_{e,(1)}^{+}$}] at (p1) {};
    \node[inner sep=0,label={[label distance=-1mm]45:$B_{e,(2)}^{+}$}] at (bl2) {};
    \node[inner sep=0,label={[label distance=-1mm]45:$B_{e,(1)}^{-}$}] at (blb1) {};
  \end{tikzpicture}
  \caption{}
  \label{subfig:incre-wl-a}
\end{subfigure}\hspace{.75em}
\begin{subfigure}[b]{.5\linewidth}
  \begin{tikzpicture}[
      line width=1pt,scale=.64,
      pin/.style={circle,fill,scale=0.7},
      hbt/.style={rectangle,fill=bole!50,draw=bole},
      btmpin/.style={pin,fill=chazblue!50,draw=chazblue},
      toppin/.style={pin,fill=chazred!50,draw=chazred},
      net/.style={line width=1pt,dashed,opacity=.6},
      seg/.style={line width=1pt,opacity=.6},
      darrow/.style={{Triangle[width=4pt,length=4pt]}-{Triangle[width=4pt,length=4pt]}}
    ]
    \node[toppin,label=-90:$p_1$] (p1) at (0,0) {};
    \node[toppin,label=90:$p_2$] (p2) at (1,3) {};
    \node[toppin,label=135:$p_3$] (p3) at (3,2) {};
    \node[btmpin,label=45:$p_4$] (p4) at (4,1) {};
    \path let\p1=(p1),\p3=(p3) in coordinate (blb2) at (\x3,\y1);
    \path let\p2=(p2),\p4=(p4) in coordinate (tr1) at (\x4,\y2);
    \path let\p2=(p2),\p3=(p3) in coordinate (tr2) at (\x3,\y2);
    \node[inner sep=0,label={[label distance=-1mm]90:$\Delta W_e^+$}] at ($(tr1)!.5!(tr2)$) {};
    \begin{pgfonlayer}{background}
    \filldraw[net,draw=chazred,fill=chazred!60] (p1) rectangle (tr2);
    \filldraw[net,draw=chazred,fill=chazred!70] (bl2) rectangle (p3);
    \draw[net] (blb2) rectangle (tr1);
    \end{pgfonlayer}
    \node[btmpin,label=180:$p_5$] (p5) at (2.5,0.5) {};
    \node[btmpin,label=135:$p_6$] (p6) at (5,-2) {};
    \path let\p4=(p4),\p5=(p5) in coordinate (tl1) at (\x5,\y4);
    \path let\p5=(p5),\p6=(p6) in coordinate (blb1) at (\x5,\y6);
    \path let\p4=(p4),\p6=(p6) in coordinate (tr3) at (\x6,\y4);
    \path let\p5=(p5),\p6=(p6) in coordinate (tr4) at (\x6,\y5);
    \node[inner sep=0,label={[label distance=-1mm]0:$\Delta W_e^-$}] at ($(tr3)!.5!(tr4)$) {};
    \begin{pgfonlayer}{background}
    \filldraw[net,draw=chazblue,fill=chazblue!60] (tl1) rectangle (p6);
    \draw[net] (p5) rectangle (tr3);
    \end{pgfonlayer}
  \end{tikzpicture}
  \caption{}
  \label{subfig:incre-wl-b}
\end{subfigure}
  \caption{Illustration of incremental computation for $\nabla_{\bm{z}}\hat{W}_{e,\text{Bi}}$ of $p_4$, other pins share the same procedure.
  \subref{subfig:incre-wl-a}~$p_4$ lies on the current bounding box $B_{e, (1)}^+$, and moving it to the bottom die makes $B_{e, (2)}^+$ become the boundary.
  \subref{subfig:incre-wl-b}~The bistratal wirelength change $\Delta W_{e, \text{Bi}} = \Delta W_e^+ + \Delta W_e^-$ can be calculated in constant time when moving $p_4$ to the bottom die.}
  \label{fig:incre-wl}
\end{figure}

\subsection{Incremental Wirelength Gradient Algorithm}
\label{subsec:wl-acl}
The depth gradient $\nabla_{\bm{z}}\hat{W}_{e,\text{Bi}}$ in~\Cref{eq:fda-grad} requires different computation mechanism compared to other weighted-average model based gradients.
We perturb the pin partition with {$\Delta z = \frac{d_z}{4}$} and check the bistratal wirelength change $\Delta W_{e, \text{Bi}}$, involving frequently updating the bounding boxes of the top and bottom nets.

Let $P_e=\{p_1,\cdots,p_l\}$ denote the set of all pins connected by net $e$.
The {vanilla approach in~\cite{TCAD23-Liao}} evaluates the $W_{e,\text{Bi}}$ after changing the partition of pin $p \in P_e$ by traversing the rest of pins $P_e \setminus \{p\}$ to check the maximum and minimum values.
The time complexity is $O(|P_e|^2)$ to finish the computation for net $e$.
However, we find that most computation in the vanilla approach is unnecessary, and the equivalent results can be calculated incrementally.

{
The key observation is that updates to the coordinates of the top-net bounding box $B_{e, (1)}^+$ with coordinates $\left\{x_{\min}^{+}, y_{\min}^{+}, x_{\max}^{+}, y_{\max}^{+}\right\}$ or the bottom-net bounding box $B_{e, (1)}^-$ with coordinates $\left\{x_{\min}^{-}, y_{\min}^{-}, x_{\max}^{-}, y_{\max}^{-}\right\}$ are necessary only if the pin being moved is located at the boundary of these boxes.
Rather than traversing all pins $P_e$ to determine new maximum and minimum values, it is sufficient to refer to the coordinates of the second outermost bounding boxes, $B_{e, (2)}^+$ and $B_{e, (2)}^-$, which represent the second-largest and second-smallest values among the pin coordinates on the top and bottom dies, respectively.
A similar approach has been adopted in~\cite{fu2024coplace}.
}

If the top pin $p$ is located on $B_{e, (1)}^+$, and is moved to the bottom die, then $B_{e, (2)}^+$ becomes the new bounding box for the top net.
If the location of $p$ on the bottom die is outside the bounding box $B_{e, (1)}^-$, we then update $B_{e, (1)}^-$ accordingly.
{
The change of bistratal wirelength $\Delta W_{e, \text{Bi}}$ can be calculated by $\Delta W_{e, \text{Bi}} = W_{e, \text{Bi}}^{p} - W_{e, \text{Bi}}$, where $W_{e, \text{Bi}}^{p}$ represents the bistratal wirelength after changing the partition of pin $p$ and $W_{e, \text{Bi}}$ represents the initial bistratal wirelength.
The calculation costs constant time for each pin.
}
Hence, the incremental algorithm computes the depth gradient $\nabla_{\bm{z}}\hat{W}_{e,\text{Bi}} = \frac{\Delta W_{e, \text{Bi}}}{\Delta z}$ in time complexity $O(|P_e|)$ for net $e$. \Cref{fig:incre-wl} illustrates the incremental computation for one pin, and the gradients for other pins can be calculated similarly.
{Compared to the vanilla approach in~\cite{TCAD23-Liao}, our incremental algorithm exhibits lower time complexity, making it more efficient.}

\begin{table}[tb!]
  \centering\footnotesize
  \caption{The statistics of ICCAD 2023 contest benchmarks~\cite{ICCAD23-Hu}. 
    $\text{RH}^{+}$ and $\text{RH}^{-}$ represent row height values of the top and bottom die, respectively. 
    $w'$ stands for the pitch size of HBTs.
    $r_{\text{MA}}$ stands for the macro area ratio, which is calculated using the technology information of the top die.}
  \label{tab:benchmark-stat}
  \resizebox{\linewidth}{!}{
    \begin{tabular}{l|rS[table-format=2.0]rS[table-format=2.0]S[table-format=3.0]S[table-format=2.0]S[table-format=1.2]}
      \toprule
      \multicolumn{1}{c|}{\textbf{Bench.}}&
      \multicolumn{1}{c}{\#Cells}&
      \multicolumn{1}{c}{\#Macros}&
      \multicolumn{1}{c}{\#Nets}&
      \multicolumn{1}{c}{$\text{RH}^{+}$}&
      \multicolumn{1}{c}{$\text{RH}^{-}$}&
      \multicolumn{1}{c}{$w'$}& 
      \multicolumn{1}{c}{$r_{\text{MA}}$}\\
      \midrule
      \texttt{case2}   &13901   &6    &19547   & 33 & 33  & 92 & 0.88\\
      \texttt{case2h1} &13901   &6    &19547   & 33 & 48  & 92 & 0.88\\
      \texttt{case2h2} &13901   &6    &19547   & 33 & 48  & 92 & 0.88\\
      \texttt{case3}   &124231  &34   &164429  & 33 & 48  & 56 & 0.71\\
      \texttt{case3h}  &124231  &34   &164429  & 36 & 48  & 58 & 0.67\\
      \texttt{case4}   &740211  &32   &758860  & 92 & 115 & 54 & 0.36\\
      \texttt{case4h}  &740211  &32   &758860  & 55 & 69  & 32 & 0.36\\
      \bottomrule
    \end{tabular}
  }
\end{table}

\begin{table}[tb!]
  \centering\footnotesize
  \caption{The statistics of RISC-V designs. The top die uses NanGate 15nm~\cite{ISPD15-nangate15} with $\text{RH}^{+} = \text{\SI{0.768}{\micro\metre}}$, and the bottom die uses NanGate 45nm~\cite{nangate45} with $\text{RH}^{-} = \text{\SI{1.4}{\micro\metre}}$. $u^{+}$ and $u^{-}$ represent the maximum utilization rate of the top and bottom die, respectively.}
  \label{tab:risc-stat}
  \resizebox{\linewidth}{!}{
    \begin{tabular}{r|rS[table-format=3.0]rS[table-format=1.2]S[table-format=1.2]S[table-format=1.2]}
      \toprule
      \multicolumn{1}{c|}{\textbf{Bench.}}&
      \multicolumn{1}{c}{\#Cells}&
      \multicolumn{1}{c}{\#Macros}&
      \multicolumn{1}{c}{\#Nets}&
      \multicolumn{1}{c}{$u^{+}$}&
      \multicolumn{1}{c}{$u^{-}$}&
      \multicolumn{1}{c}{$r_{\text{MA}}$}\\
      \midrule
      \texttt{tinyRocket}   &24647   &2    &26085   & 0.50 & 0.60   & 0.07\\
      \texttt{SweRV}   &87587   &28    &91903   & 0.70 & 0.80   & 0.81\\
      \texttt{Ariane}   &145684   &132    &157129   & 0.80 & 0.90   & 0.67\\
      \texttt{BlackParrot}   &273187   &24    &265585   & 0.55 & 0.65   & 0.55\\
      \bottomrule
    \end{tabular}
  }
\end{table}

\begin{table}[tb!]
  \footnotesize
  \centering
  \caption{The \emph{official} raw score comparison with top-3 winners provided by ICCAD 2023 contest. The raw score = \text{HPWL} + $\beta$\#\text{HBTs}. $\beta$ is 10 for all the cases.}
  \label{tab:main-result}
  \begin{threeparttable}
    \begin{tabular}{l|*{2}{S[table-format=10.0]S[table-format=10.0]}}
      \toprule
      \multicolumn{1}{c|}{\multirow{1}{*}{\textbf{Bench.}}} & \multicolumn{1}{c}{1st Place} & \multicolumn{1}{c}{2nd Place} &  \multicolumn{1}{c}{3rd Place} & \multicolumn{1}{c}{Ours}\\
      \midrule
      \texttt{case2}   & 16506066 & 16287082 & 16559126 & \ubold 15635352\\
      \texttt{case2h1} & 18123044 & 19055977 & 21180946 & \ubold 16569703\\
      \texttt{case2h2} & 18124483 & 19202109 & 21664974 & \ubold 16820960\\
      \texttt{case3}   & 98928220 & 105647967 & 116317085 & \ubold 98206238\\
      \texttt{case3h}  & 122459408 & 120820762 & 117889705 & \ubold 108166770\\
      \texttt{case4}   & 1047716115 & 1110850494 & 1131599485 & \ubold 1037676163\\
      \texttt{case4h}  & 656528147 & 682231267 & 703663946 & \ubold 635259476\\
      \midrule
      \multicolumn{1}{c|}{Average} & {1.059} & {1.096} & {1.157} & {\ubold 1.000} \\
      \bottomrule
    \end{tabular}
  \end{threeparttable}
\end{table}

\begin{table*}[htb!]
  \footnotesize
  \centering
  \caption{Score decomposition compared to the top-3 winners {on the ICCAD 2023 contest benchmark}. \textbf{RT} (s) stands for the total runtime. {All the baselines are evaluated on our machine with 8 CPU threads. Our placer is evaluated  both on the CPU with 8 threads and on the GPU.}}
  \label{tab:decomposed-result}
  \begin{threeparttable}
    \resizebox{\linewidth}{!}{
    \begin{tabular}{l*{2}{|S[table-format=10.0]S[table-format=6.0]S[table-format=4.0]|S[table-format=10.0]S[table-format=6.0]S[table-format=4.0]}|S[table-format=10.0]S[table-format=6.0]S[table-format=3.0]}
      \toprule
      \multicolumn{1}{c|}{\multirow{2}{*}{\textbf{Bench.}}} & \multicolumn{3}{c|}{1st Place} & \multicolumn{3}{c|}{2nd Place} &  \multicolumn{3}{c|}{3rd Place} & \multicolumn{3}{c|}{Ours-CPU} & \multicolumn{3}{c}{Ours-GPU}\\
      & {HPWL} & {\#HBTs} & RT & {HPWL} & {\#HBTs} & RT & {HPWL} & {\#HBTs} & RT & { HPWL} & { \#HBTs} & { RT} & { HPWL} & { \#HBTs} & { RT} \\
      \midrule
      \texttt{case2}   & 16490836 & 1523 & 67 & 16277152 & 993 & 32 & 16537596 & 2153 & 144 & \ubold 15528810 & 1128 & 76  & 15622062 & 1329 & 38\\
      \texttt{case2h1} & 18121844 & 120 & 39 & 19047767 & 821 & 43 & 21156596 & 2435 & 160 &  16708363 & 1135 & 80 & \ubold 16556213 & 1349 & 35\\
      \texttt{case2h2} & 18123283 & 120 & 42 & 19193899 & 821 & 41 & 21640714 & 2426 & 162 & \ubold 16748148 & 1091 & 80 &  16807840 & 1312 & 36\\
      \texttt{case3}   & 98706330 & 22189 & 534 & 105386847 & 26112 & 104 & 116022515 & 29457 & 602 & \ubold 97281904 & 10704 & 236 &  98081778 & 12446 & 92\\
      \texttt{case3h}  & 122271798 & 18761 & 262 & 120770382 & 5038 & 104 & 117633295 & 25641 & 612 &  109386719 & 13224 & 239 & \ubold 108028790 & 13798 & 86\\
      \texttt{case4}   & 1046106185 & 160993 & 3605 & 1108969124 & 188137 & 615 & 1130211865 & 138762 & 5309 &  1040202500 & 132109 & 3070 & \ubold 1036364973 & 131119 & 335\\
      \texttt{case4h}  & 654962287 & 156586 & 1567 & 680554407 & 167686 & 592 & 702244786 & 141916 & 4492 &  634654510 & 133734 & 2640 & \ubold 633920946 & 133853 & 361\\
      \midrule
      \multicolumn{1}{c|}{Average} & {1.058} & {0.981} & {4.000} & {1.096} & {1.019} & {1.289} & {1.156} & {1.660} & {7.830} & { 1.000} & {0.907} & {4.047} & {\ubold 1.000} & {1.000} & {1.000} \\
      \bottomrule
    \end{tabular}
    }
  \end{threeparttable}
\end{table*}

\begin{table*}[tb!]
  \footnotesize
  \centering
  \caption{{ Experimental results on modern RISC-V designs. HPWL is measured in \SI{}{\micro\metre}. The baseline is evaluated on our machine with 8 CPU threads. Our placer is evaluated  both on the CPU with 8 threads and on the GPU.}}
  \label{tab:risc-result}
  \begin{threeparttable}
    \begin{tabular}{r*{1}{|S[table-format=7.0]S[table-format=7.0]S[table-format=5.0]S[table-format=4.0]}*{2}{|S[table-format=7.0]S[table-format=7.0]S[table-format=5.0]S[table-format=3.0]}}
      \toprule
      \multicolumn{1}{c|}{\multirow{2}{*}{ \textbf{Bench.}}} & \multicolumn{4}{c|}{ 1st Place} & \multicolumn{4}{c|}{ Ours-CPU} & \multicolumn{4}{c}{ Ours-GPU}\\
      & { Score} & { HPWL} & { \#HBTs} & { RT}  & { Score} & { HPWL} & { \#HBTs} & { RT} & { Score} & { HPWL} & { \#HBTs} & { RT} \\
      \midrule
      {\texttt{tinyRocket}}   & 206290 & 145830 & 6046 & 356 & 182559 & 130749 & 5181 & 104  & \ubold 181571 & \ubold 130631 & 5094 & 42\\
      {\texttt{SweRV}} & 1020414 & 981634 & 3878 & 2208 & 924932 & 805952 & 11898 & 341 & \ubold 922004 &  \ubold 800464 & 12154 & 108\\
      {\texttt{Ariane}} & 1828336 & 1477246 & 35109 & 1009 & 1328717 & 1184607 & 14411 & 397 & \ubold 1311346 & \ubold 1174076 & 13727 & 142\\
      {\texttt{BlackParrot}}  & NA  & NA & NA & NA & \ubold 1586522 & \ubold 1488882 & 9764 & 640 & 1613662 & 1515872 & 9779 & 156\\
      \midrule
      \multicolumn{1}{c|}{Average} & {1.212} & {1.200} & {1.355} & {12.009} & {1.001} & { 1.000} & {1.011} & {3.133} & {\ubold 1.000} & {\ubold 1.000} & {1.000} & {1.000} \\
      \bottomrule
    \end{tabular}
  \end{threeparttable}
\end{table*}

\newenvironment*{tikzhackaxes3d}[2][]{
  \begin{tikzpicture}[node distance=4.25pt,align=left]
    \scriptsize
    \node[inner sep=0] (fig) {\includegraphics[#1]{#2}};
    \node[anchor=north,inner sep=.2mm,fill=slategray!10] at (fig.south)
    \bgroup\ignorespaces%
    \setlength\tabcolsep{3pt}
}{
    \unskip\egroup;
  \end{tikzpicture}
}
\newenvironment*{tikzhackaxes2d}[2][]{
  \begin{tikzpicture}[node distance=11.8pt]
    \footnotesize
    \node[inner sep=0] (fig) {\includegraphics[#1]{#2}};
    \node[fill=slategray!10,anchor=north] (fig tag) at (fig.south)
    \bgroup\ignorespaces
}{
    \unskip\egroup;
    \node[anchor=north] at ($(fig tag.north)+(0,-0.25)$) {};
    \node[anchor=south] at (fig.north) {};
  \end{tikzpicture}
}

\begin{figure*}[htb]
  \centering
  \begin{tabular}{cccc}
    \begin{tikzhackaxes3d}[width=.22\linewidth]{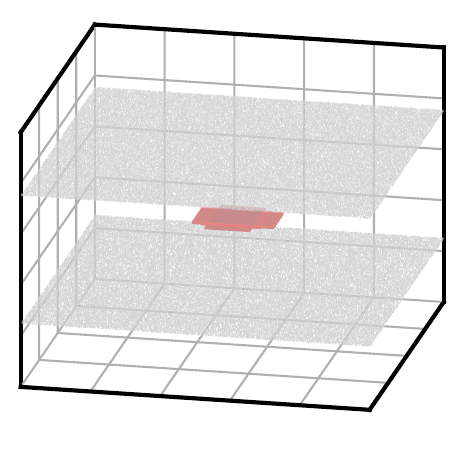}%
      \begin{tabular}{cccc}
          \bfseries Iter. & \bfseries WL (\power{7}) & \bfseries \#HBTs & \bfseries OVFL\\
          0 & 3.60 & 88507 & 1.00\\
      \end{tabular}
    \end{tikzhackaxes3d}&
    \begin{tikzhackaxes3d}[width=.22\linewidth]{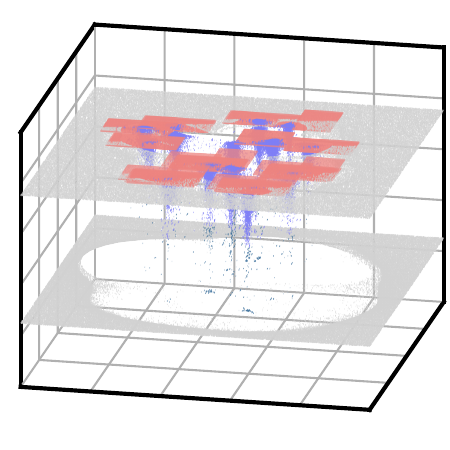}%
      \begin{tabular}{cccc}
          \bfseries Iter. & \bfseries WL (\power{7}) & \bfseries \#HBTs & \bfseries OVFL\\
          1000 & 4.49 & 1254 & 0.89\\
      \end{tabular}
    \end{tikzhackaxes3d}&
    \begin{tikzhackaxes3d}[width=.22\linewidth]{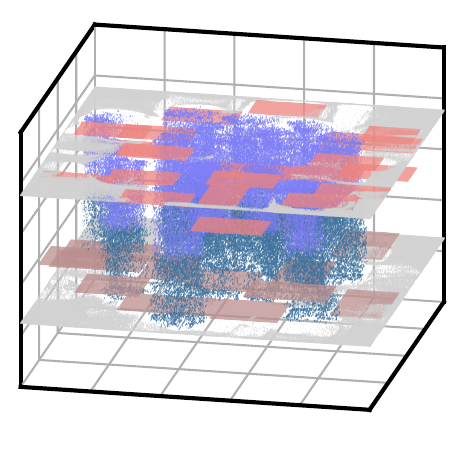}%
      \begin{tabular}{cccc}
          \bfseries Iter. & \bfseries WL (\power{7}) & \bfseries \#HBTs & \bfseries OVFL\\
          1800 & 10.30 & 30961 & 0.34\\
      \end{tabular}
    \end{tikzhackaxes3d}&
    \begin{tikzhackaxes3d}[width=.22\linewidth]{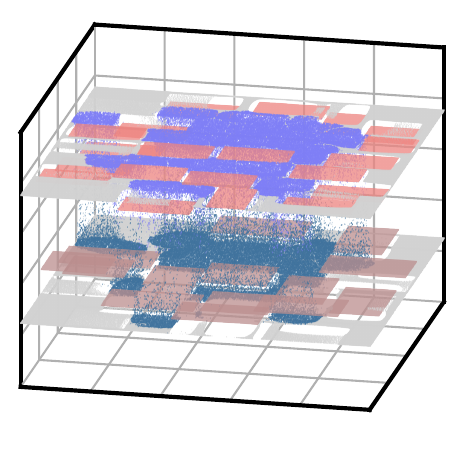}%
      \begin{tabular}{cccc}
          \bfseries Iter. & \bfseries WL (\power{7}) & \bfseries \#HBTs & \bfseries OVFL\\
          2142 & 10.51 & 13798 & 0.07\\
      \end{tabular}
    \end{tikzhackaxes3d}
  \end{tabular}
  \caption{
    The 3D mixed-size global placement process on \texttt{case3h} with heterogeneous technology nodes. 
    Macros and standard cells spread at the same speed at the early global placement stage, leading to an optimized macro partitioning subsequently. 
    The convergent placement solution with overflow 0.07 finds a clear instance partitioning.
  }
  \label{fig:case3h-gp-iter}
\end{figure*}

\begin{figure*}[htb]
  \centering
  \begin{tabular}{cccc}
    \makebox[.23\linewidth][c]{
      \begin{tikzpicture}
        \footnotesize
        \node[inner sep=0] (fig) {%
          \includegraphics[width=.22\linewidth]{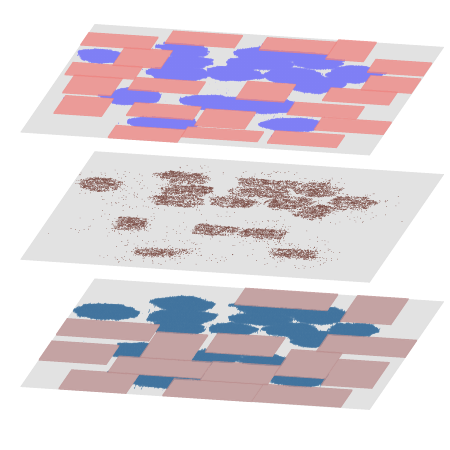}};%
          \node[anchor=south] at (fig.north) {};
      \end{tikzpicture}
    }&
    \begin{tikzhackaxes2d}[width=.22\linewidth]{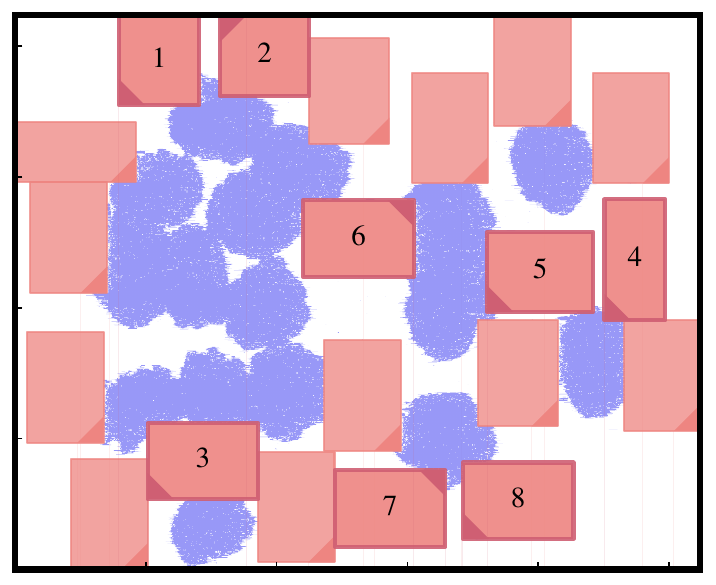}%
      \footnotesize\bfseries top die
    \end{tikzhackaxes2d}&
    \begin{tikzhackaxes2d}[width=.22\linewidth]{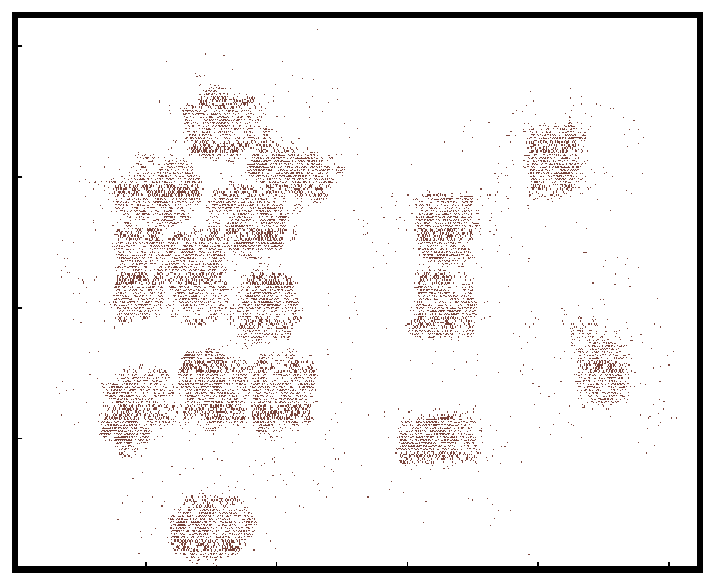}%
      \footnotesize\bfseries HBTs
    \end{tikzhackaxes2d}&
    \begin{tikzhackaxes2d}[width=.22\linewidth]{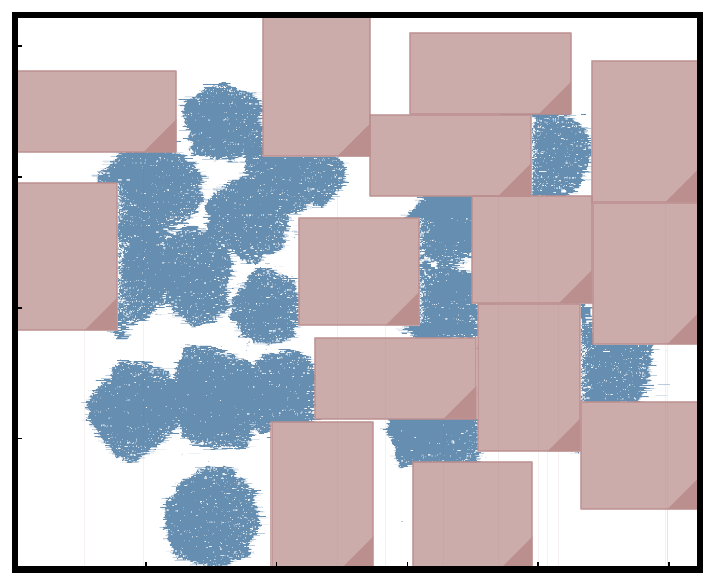}%
      \footnotesize\bfseries bottom die
    \end{tikzhackaxes2d}%
  \end{tabular}
  \caption{
    The final layout of \texttt{case3h} after the detailed placement. 
    Our MILP finds the optimized macro rotations, consequently improving wirelength. 
    The 8 rotated macros are marked with numbers.
    The hybrid bonding terminals are sparsely placed to connect the instances on different dies.
  }
  \label{fig:case3h-final-sol}
\end{figure*}

\section{Experimental Results}
\label{sec:results}
\subsection{Experimental Setup}
\label{subsec:exp-setup}
We conducted experiments on the ICCAD 2023 contest benchmarks~\cite{ICCAD23-Hu} { and open-source RISC-V designs}. 
The detailed design statistics { of ICCAD 2023 contest benchmark} are shown in~\Cref{tab:benchmark-stat}. 
Top and bottom maximum utilization rate is 80\%, and the HBT cost $\beta$ is 10 for { these} designs.
The contest evaluates the \emph{raw score} = HPWL + $\beta$\#HBTs with a runtime factor.
Most designs adopt heterogeneous technology nodes with a large macro area ratio $r_{\text{MA}}$, bringing a significant challenge to optimizing the D2D wirelength.
{ 
Additionally, we evaluated our placer on four modern RISC-V designs including \texttt{tinyRocket}~\cite{tinyRocket}, \texttt{SweRV}~\cite{swerv}, \texttt{Ariane}~\cite{Ariane}, and \texttt{BlackParrot}~\cite{BlackParrot}.
The RTL designs were synthesized using Yosys in OpenROAD project~\cite{DAC19-OpenRoad}.
The heterogeneous F2F stacking is set as NanGate 15nm~\cite{ISPD15-nangate15} on the top die and NanGate 45nm~\cite{nangate45} on the bottom die.
The HBT pitch size $w'$ is \SI{1}{\micro\metre}, and the HBT cost $\beta$ is 10 for these designs.
The detailed design statistics are shown in~\Cref{tab:risc-stat}.
}

We implemented the proposed 3D mixed-size placement framework in \texttt{C++} and \texttt{CUDA} based on the open-source placer {DREAMPlace}~\cite{DAC19-Lin}. 
And we used \texttt{Gurobi}~\cite{Gurobi} as the MILP solver.
We set the $z$-bin size as $d_{b} = \frac{w_{b} + h_{b}}{2}$, and the region $\Omega$ depth is $d_{z}=N_{z}d_{b}$.
We empirically set the HBT penalty factor as $\alpha=\alpha_0\frac{d_x\eta^2}{d_z}\log\left(90\beta{\eta}-1\right)$ where $\alpha_0=\text{3.5}\times\text{10\textsuperscript{$-$3}}$ and $\eta=\frac{2w'}{\text{RH}^{+}+\text{RH}^{-}}$, { considering the relationship between number of HBTs and design statistics.}
All the experiments were performed on a Linux machine with 20-core Intel Xeon Silver 4210R CPU (2.40GHz), 1 GeForce RTX 3090Ti GPU, and 24GB RAM.
We compared our framework with the state-of-the-art (SOTA) placers from top-3 teams in the ICCAD 2023 contest~\cite{ICCAD23-Hu}, and the reported results were evaluated using the official evaluator provided by the contest.
{
We obtained the executables from the contest winners and ran them on our machine with 8 CPU threads following the contest setting.
Our placer was evaluated both on the CPU with 8 threads and on the GPU.
}

\subsection{Comparison with SOTA Placers}
\label{subsec:comparison-stoa}
\Cref{tab:main-result} shows the official raw score of top-3 teams and our framework on the contest benchmarks.
We also compared the detailed score decomposition including D2D HPWL and HBT number, and reported the runtime of each case with the baselines in~\Cref{tab:decomposed-result}.
Our analytical 3D placement framework consistently obtained the best results for all the cases, as shown in~\Cref{tab:main-result}, demonstrating the significant advantage of our 3D placement paradigm with the dedicated density model and  bistratal wirelength model.
Compared to the top-3 teams, our placer achieved {5.9\%, 9.6\%, and 15.7\%} better score on average, respectively.
The score is dominated by the wirelength due to the small HBT cost.
Our placer obtained better wirelength than the baselines on all the cases, using a similar number of HBTs, as shown in~\Cref{tab:decomposed-result}.
{
When running on a CPU, our placer shows similar runtime with the first place and achieves 2.1$\times$ speedup over the third place.
And our framework demonstrates better scalability, achieving up to {2.3$\times$} and {2.6$\times$} speedup over the first and third places, respectively, on the large cases.
While the second place is 2.9$\times$ faster than our placer, it is limited to 2D placement engine and yields lower quality results.
Additionally, our proposed algorithms are suitable for GPU acceleration.
}
Leveraging the adaptive 3D density accumulation and incremental wirelength gradient algorithms, our GPU-accelerated placer { shows significant runtime improvements compared to the baselines.}
Specifically, it achieves {4.0$\times$ and 7.8$\times$} speedup over the first and third places, respectively, and up to {1.8$\times$} speedup over the second place on the large cases.

{
We also evaluated our framework on four modern RISC-V designs, with experimental results shown in~\Cref{tab:risc-result}.
The second and third places obtained very low-quality results and failed to generate legal solutions for more than two designs, so their results have been omitted.
The first place failed for \texttt{BlackParrot} because of diverged 3D global placement and subsequent macro legalization error.
In contrast, our placer obtained legal placement solution across all tested designs.
Compared to the first place, our placer not only reduced wirelength by a significant 20\% but also achieved a 4.1$\times$ speedup on CPU and a remarkable 12.0$\times$ speedup on GPU.
}

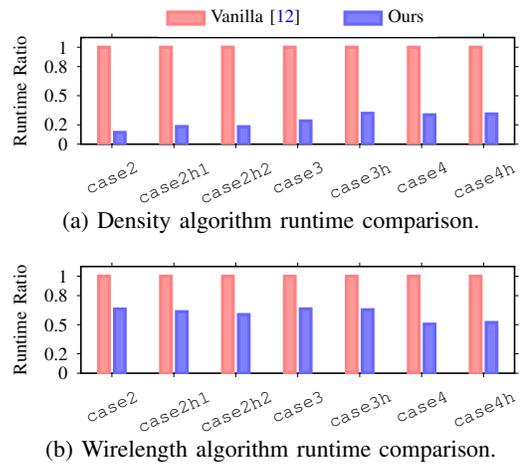
\begin{figure}[tb!]
  \centering
  \begin{subfigure}{\linewidth}
  \centering
  \begin{tikzpicture}[font=\scriptsize]
    \begin{axis}[
        ybar,
        scale only axis,clip=true,
        xmin = 0.5, xmax = 7.5,
        ymin = 0, ymax = 1.1,
        width=0.65\linewidth,
        height=0.16\linewidth,
        axis line style={line width=.6pt},
        legend cell align=left,
        tick align=outside,
        bar width = 4pt,
        every major tick/.append style={ %
          major tick length=1.5pt, black},
        xlabel style={font=\scriptsize},
        x tick label style={
            rotate=25, font=\scriptsize},
        xticklabels={\texttt{case2}, \texttt{case2h1}, \texttt{case2h2}, \texttt{case3}, \texttt{case3h}, \texttt{case4}, \texttt{case4h}},
        x tick label style={
          /pgf/number format/assume math mode, font=\scriptsize},
        y tick label style={
          /pgf/number format/assume math mode, font=\scriptsize},
        ytick={0.0, 0.2, 0.5, 0.8, 1.0},
        xtick={1, 2, 3, 4, 5, 6, 7},
        ylabel={Runtime Ratio},
        legend image post style={scale=0.8},
        legend style={
            draw=none,
            at={(0.5,1.)},
            anchor=south,
            legend columns=2,
            /tikz/every even column/.append style={column sep=0.5cm}
        },
        every axis plot/.append style={smooth,tension=0.1,line width=1.2pt}
      ]
      \addplot[ybar, fill=red!40, draw=red!50, area legend] table [x index=0,y index=1,col sep=comma] {data/density-vanilla.csv};
      \addplot[ybar, fill=blue!50!white, draw=blue!60!white, area legend] table [x index=0,y index=1,col sep=comma] {data/density-ours.csv};
      \legend{Vanilla~\cite{TCAD23-Liao}, Ours}
    \end{axis}
  \end{tikzpicture}
  \vspace*{-3mm}
  \caption{Density algorithm runtime comparison.}
  \vspace*{2mm}
  \label{subfig:acc-density}
\end{subfigure} \\
\begin{subfigure}{\linewidth}
  \centering
  \begin{tikzpicture}[font=\scriptsize]
    \begin{axis}[
        ybar,
        scale only axis,clip=true,
        xmin = 0.5, xmax = 7.5,
        ymin = 0, ymax = 1.1,
        width=0.65\linewidth,
        height=0.16\linewidth,
        axis line style={line width=.6pt},
        legend cell align=left,
        tick align=outside,
        bar width = 4pt,
        every major tick/.append style={ %
          major tick length=1.5pt, black},
        xlabel style={font=\scriptsize},
        x tick label style={
            rotate=25, font=\scriptsize},
        xticklabels={\texttt{case2}, \texttt{case2h1}, \texttt{case2h2}, \texttt{case3}, \texttt{case3h}, \texttt{case4}, \texttt{case4h}},
        x tick label style={
          /pgf/number format/assume math mode, font=\scriptsize},
        y tick label style={
          /pgf/number format/assume math mode, font=\scriptsize},
        ytick={0.0, 0.2, 0.5, 0.8, 1.0},
        xtick={1, 2, 3, 4, 5, 6, 7},
        ylabel={Runtime Ratio},
        legend image post style={scale=0.8},
        legend style={
            draw=none,
            at={(0.5,1.)},
            anchor=south,
            legend columns=2,
            /tikz/every even column/.append style={column sep=0.5cm}
        },
        every axis plot/.append style={smooth,tension=0.1,line width=1.2pt}
      ]
      \addplot[ybar, fill=red!40, draw=red!50, area legend] table [x index=0,y index=1,col sep=comma] {data/wl-vanilla.csv};
      \addplot[ybar, fill=blue!50!white, draw=blue!60!white, area legend] table [x index=0,y index=1,col sep=comma] {data/wl-ours.csv};
    \end{axis}
  \end{tikzpicture}
  \vspace*{-3mm}
  \caption{Wirelength algorithm runtime comparison.}
  \label{subfig:acc-wl}
\end{subfigure}
  \caption{Runtime comparison of \subref{subfig:acc-density} density and \subref{subfig:acc-wl} wirelength algorithms between vanilla approach in~\cite{TCAD23-Liao} and ours on GPU.}
  \label{fig:acc-density}
\end{figure}

\subsection{3D Mixed-Size Placement Analysis}
The 3D mixed-size global placement plays a dominant role in our framework, which optimizes the D2D wirelength while explicitly considering instance partitioning, visualized in~\Cref{fig:case3h-gp-iter}.
Fillers, standard cells on the top die, and standard cells on the bottom die are denoted by gray, purple, and blue rectangles.
The macros are colored in red on the top die and colored in brown on the bottom die.
The instance depth is omitted for clear visualization.
All the standard cells and macros are randomly initialized around the center of the region from a normal distribution.
During the global placement, the bistratal wirelength model effectively optimizes instance locations in the 3D solution space.
The proposed 3D preconditioner allows macros and standard cells to spread at the same speed, leading to an optimized macro partitioning at a later stage.
The customized density model finally drives all the instances to exactly two dies.

{
\subsection{Acceleration of Density and Wirelength Algorithms}
\label{subsec:acc-density-wl}
We further investigate the efficiency of our proposed density and wirelength algorithms.
\Cref{subfig:acc-density} compares our adaptive 3D density accumulation with the instance-parallel approach~\cite{TCAD23-Liao}.
Our approach achieves 4.7$\times$ speedup by exploiting the abundant bin parallelism for macros, avoiding the load balancing issue.

\Cref{subfig:acc-wl} compares our incremental wirelength gradient algorithm with the vanilla approach in~\cite{TCAD23-Liao}.
Our algorithm reduces the time complexity from $O(|P_e|^2)$ to $O(|P_e|)$ for calculating the $z$-gradient, resulting in 1.7$\times$ speedup.
}

\begin{figure}[tb!]
  \begin{subfigure}[b]{.48\linewidth}
  \centering
  \begin{tikzpicture}[font=\scriptsize,trim axis left]
    \begin{axis}[
        scale only axis,clip=true,
        at={(0,0)},
        width=0.8\linewidth,
        height=0.56\linewidth,
        axis line style={line width=.6pt},
        legend cell align=left,
        tick align=outside,
        every major tick/.append style={ %
          major tick length=1.5pt, black},
        xlabel style={font=\scriptsize},
        x tick label style={
          /pgf/number format/1000 sep={}},
        xlabel={},
        ymax=2.1,
        x tick label style={
          /pgf/number format/assume math mode, font=\scriptsize},
        y tick label style={
          /pgf/number format/assume math mode, font=\scriptsize},
        legend style={at={(rel axis cs:.0,1)},
          draw=none,anchor=north west,fill=none,font=\scriptsize},
        every axis plot/.append style={smooth,tension=0.1,line width=1.2pt}
      ]
      \addplot[color=red!50!white] table [x index=0,y index=1,col sep=comma] {data/case4-hpwl-q-p.csv};
      \addplot[color=green!50!white] table [x index=0,y index=1,col sep=comma] {data/case4-hpwl-q.csv};
      \addplot[color=blue!50!white] table [x index=0,y index=1,col sep=comma] {data/case4-hpwl.csv};
    \end{axis}
    \node[anchor=south west,align=center,inner sep=2mm] at (rel axis cs:0.02,1.05) {$\cdot$10\textsuperscript{9}};
  \end{tikzpicture}
  \caption{Wirelength curve.}
\end{subfigure}\hspace{.75em}
\begin{subfigure}[b]{.48\linewidth}
  \centering
  \begin{tikzpicture}[font=\scriptsize,trim axis left]
    \begin{axis}[
        scale only axis,clip=true,
        at={(0,0)},
        width=0.8\linewidth,
        height=0.56\linewidth,
        axis line style={line width=.6pt},
        legend cell align=left,
        tick align=outside,
        every major tick/.append style={ %
          major tick length=1.5pt, black},
        xlabel style={font=\scriptsize},
        x tick label style={
          /pgf/number format/1000 sep={}},
        xlabel={},
        x tick label style={
          /pgf/number format/assume math mode, font=\scriptsize},
        y tick label style={
          /pgf/number format/assume math mode, font=\scriptsize},
        legend style={at={(rel axis cs:0,0)},
          draw=none,anchor=south west,fill=none,font=\scriptsize},
        every axis plot/.append style={smooth,tension=0.1,line width=1.2pt}
      ]
      \addplot[color=red!50!white] table [x index=0,y index=1,col sep=comma] {data/case4-macro-overflow-q-p-scale.csv};
      \addplot[color=green!50!white] table [x index=0,y index=1,col sep=comma] {data/case4-macro-overflow-q-scale.csv};
      \addplot[color=blue!50!white] table [x index=0,y index=1,col sep=comma] {data/case4-macro-overflow-scale.csv};
      \legend{ePlace, ePlace3D, Ours}
    \end{axis}
    \node[anchor=south west,align=center,inner sep=2mm] at (rel axis cs:0.02,1.043) {$\cdot$10\textsuperscript{-2}};
  \end{tikzpicture}
  \caption{Macro Density Overflow.}
\end{subfigure}
  \caption{Wirelength and macro density overflow curves over global placement iterations of different preconditioners on \texttt{case4}. 
  The macro density overflow is calculated by the macro density map and the target density.}
  \label{fig:precond}
\end{figure}
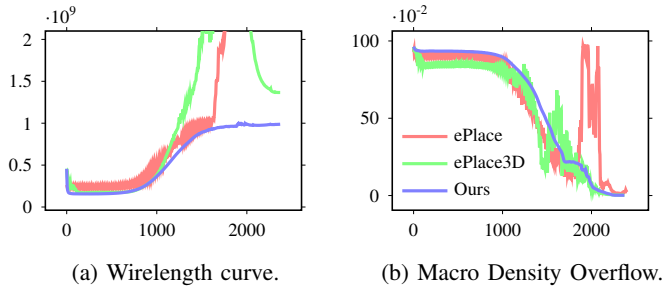

\subsection{Ablation Study on 3D Mixed-Size Preconditioning}
\label{subsec:exp-precond}
The 3D mixed-size placement is a highly nonlinear, non-convex, and ill-conditioned problem.
The heterogeneous scenarios make the problem even more complex.
The preconditioner should handle the large topological and physical difference between macros and standard cells.
Replacing our proposed preconditioner with previous approaches adopted in { ePlace}~\cite{TODAES2015-Lu-ePlace, PLACE-TCAD2015-Lu} and {ePlace3D}~\cite{ISPD16-Lu}, the 3D global placement will diverge or obtain very low-quality results with wirelength increased by 30\% on the ICCAD 2023 contest benchmarks~\cite{ICCAD23-Hu}.
\Cref{fig:precond} shows the effect of our 3D mixed-size preconditioner on \texttt{case4}.
And the trend for other designs is similar.
The previous approaches~\cite{TODAES2015-Lu-ePlace,ISPD16-Lu, PLACE-TCAD2015-Lu} fail to stabilize the optimization of macro locations for the whole process, causing macro density overflow oscillation and wirelength divergence.
In contrast, our preconditioner makes the standard cells and macros equalized in the optimizer's perspective, enabling stable optimization.

\begin{table}[tb!]
  \footnotesize
  \centering
  \caption{The raw score and runtime results of our approach without and with MILP macro rotation. \#Rot stands for the number of rotated macros.}
  \label{tab:rotation-result}
  \begin{threeparttable}
    \begin{tabular}{l|*{1}{S[table-format=10.0]S[table-format=3.0]|S[table-format=10.0]S[table-format=3.0]S[table-format=1.0]}}
      \toprule
      \multicolumn{1}{c|}{\multirow{2}{*}{\textbf{Bench.}}} & \multicolumn{2}{c|}{w/o. Rotation} & \multicolumn{3}{c}{w. Rotation}\\
      & {Score} & RT & {Score} & RT & {\#Rot} \\
      \midrule
      \texttt{case2}     & 15635352 & 38 & 15635352 & 38 & 0\\
      \texttt{case2h1}   & 16569703 & 35 & 16569703 & 35 & 0\\
      \texttt{case2h2}   & 16820960 & 36 & 16820960 & 36 & 0\\
      \texttt{case3}     & 100227409 & 91 & 98206238 & 92 & 6\\
      \texttt{case3h}    & 111062583 & 88 & 108166770 & 86 & 8\\
      \texttt{case4}     & 1058535164 & 336 & 1037676163 & 335 & 8\\
      \texttt{case4h}    & 645574820 & 346 & 635259476 & 361 & 8\\
      \midrule
      \multicolumn{1}{c|}{Average} & {1.012} & {0.996} & {1.000} & {1.000} & {-}\\
      \bottomrule
    \end{tabular}
  \end{threeparttable}
\end{table}

\subsection{Ablation Study on MILP Macro Rotation}
\label{subsec:exp-rotation}
Our MILP utilizes the physical information of the initial 3D placement solution, finding the macro rotations with optimal wirelength.
The effect of MILP macro rotation is shown in~\Cref{tab:rotation-result}. 
Except for the cases with few extremely large macros, our MILP finds the macro rotations leading to better wirelength, achieving on average 1.2\% wirelength improvement.
Since we only consider the nets connecting to macros, the runtime overhead is negligible, and it takes less than 1s for all the cases.
{ Compared to other cases, we observe a larger difference of global placement iterations for \texttt{case4h}, leading to a larger runtime difference.}
The final placement solution for \texttt{case3h} with macro rotation is illustrated in~\Cref{fig:case3h-final-sol}.

\begin{table}[tb!]
  \footnotesize
  \centering
  \caption{{ The raw score and runtime results for multi-die 2D global placement flow and 3D mixed-size global placement flow. $r_{\text{MA}}$ stands for the macro area ratio.}}
  \label{tab:place-flow}
  \begin{threeparttable}
    \begin{tabular}{l|*{1}{S[table-format=1.2]|S[table-format=10.0]S[table-format=3.0]|S[table-format=10.0]S[table-format=3.0]}}
      \toprule
      \multicolumn{1}{c|}{\multirow{2}{*}{\textbf{Bench.}}} & \multicolumn{1}{c|}{\multirow{2}{*}{$r_{\text{MA}}$}}& \multicolumn{2}{c|}{ multi-die 2D} & \multicolumn{2}{c}{ 3D mixed-size} \\
      & & { Score} & { RT} & { Score} & { RT} \\
      \midrule
      \texttt{case2}     & 0.88 & \ubold 15635352 & 38 & 17081257 & 33\\
      \texttt{case2h1}   & 0.88 & \ubold 16569703 & 35 & 17413812 & 36\\
      \texttt{case2h2}   & 0.88 & \ubold 16820960 & 36 & 17701896 & 37\\
      \texttt{case3}     & 0.71 & \ubold 98206238 & 92 & 101230278 & 118 \\
      \texttt{case3h}    & 0.67 & \ubold 108166770 & 86 & 112539329 & 100 \\
      \texttt{case4}     & 0.36 & 1064731451 & 299 & \ubold 1037676163 & 335 \\
      \texttt{case4h}    & 0.36 & 662128786 & 300 & \ubold 635259476 & 361 \\
      \midrule
      \multicolumn{1}{c|}{Average} & {-} & {0.974} & {0.923} & {1.000} & {1.000}\\
      \bottomrule
    \end{tabular}
  \end{threeparttable}
\end{table}

{
\subsection{Ablation Study on Placement Flow}
\label{subsec:exp-place-flow}
Our framework adopts multi-die 2D global placement for designs with a macro area ratio exceeding 50\%, while 3D mixed-size global placement is utilized for other designs.
\Cref{tab:place-flow} presents the experimental results comparing these two placement flows.
We observe that multi-die 2D global placement achieves 5.3\% better score than 3D mixed-size global placement for designs with a large macro footprint.
This improvement is attributed to the removal of macro density obstacle in the $z$-direction, which results in better macro placement results.
Conversely, 3D mixed-size global placement excels in optimizing standard cell locations, obtaining 3.4\% better score compared to its counterpart.
}

\section{Conclusion}
\label{sec:conclusion}
This paper proposes a new analytical 3D mixed-size placement framework with full-scale GPU acceleration, leveraging dedicated density and wirelength algorithms, for heterogeneous face-to-face (F2F) bonded 3D ICs.
Our customized density model and bistratal wirelength model, incorporating a novel 3D preconditioner, enable stable optimization for macros and standard cells in a 3D solution space.
We further propose an MILP formulation for macro rotation to optimize the wirelength.
Experimental results on ICCAD 2023 contest benchmarks demonstrate that our framework significantly surpasses the first-place winner by 5.9\% on the quality of results with {4.0}$\times$ runtime speedup.
{
Additional experiments on modern RISC-V designs further validate the generalizability and superior performance of our framework.
}

{
  \bibliographystyle{IEEEtran}
  \bibliography{ref/Top-sim,ref/All}
}

\appendix
\label{appendix}
We use notation $\prod_{\mathrm{cyc}}$ to represent multiplication over all three dimenstions, \emph{e.g.}, $\prod_{\mathrm{cyc}}f(x)=f(x)f(y)f(z)$ for any well-defined function $f$. Function $\mu(\cdot)$ is a measure defined for any measurable regions. In our three-dimensional regions, $\mu(\cdot)$ stands for the volume estimator $\vol(\cdot)$. To prove~\Cref{thm:basis-prefix-sum-3d}, we first present a lemma.
\begin{mylemma}
  \label{lem:prefix-sum}
  Denote $D^{(x,y,z)}=[x,d_x]\times[y,d_y]\times[z,d_z]$ for any $(x,y,z)\in \Omega$.
  Then $\varphi(\mathcal{A}^{(x,y,z)})_b=\frac{\mu(D^{(x,y,z)}\cap b)}{\mu(b)}$ holds for any bin $b\in B$.
\end{mylemma}
\begin{proof}
  The total number of bins is $|B|=N_xN_yN_z$. Consider the normalized coordinate $(\hat{x},\hat{y},\hat{z})=(\frac{x}{w_b},\frac{y}{h_b},\frac{z}{d_b})$ and arbitrary bin $b$ with index $(i,j,k)$. Clearly, we have $b=b_x\times b_y\times b_z$ where $b_x=[(i-1)w_b,iw_b]$, $b_y=[(j-1)h_b,jh_b]$ and $b_z=[(k-1)d_b,kd_b]$. Therefore, it must be true that
  \begin{equation}
    \frac{\mu(D^{(x,y,z)}\cap b)}{\mu(b)}=\frac{1}{w_bh_bd_b}\prod_{\mathrm{cyc}}\mu([x,d_x]\cap b_x).
    \label{eq:overlap-decomposition}
  \end{equation}
  Consider the $x$ dimension only as the other two dimensions are symmetric. If $i<\lceil\hat{x}\rceil$, we have $i<\hat{x}$ and then $iw_b<x$, which means $\mu([x,d_x]\cap b_x)=0$. If $i>\lceil\hat{x}\rceil$, we have $i-1\geq\hat{x}$ and then $(i-1)w_b\geq x$, which means $\mu([x,d_x]\cap b_x)=\mu(b_x)=w_b$. If $i=\lceil\hat{x}\rceil$, we have $x\in b_x$ and then $\mu([x,d_x]\cap b_x)=iw_b-x$. Hence, we can summarize that
  \begin{equation}
    \frac{\mu([x,d_x]\cap b_x)}{w_b}=\left\{
    \begin{array}{ll}
      0,&\text{if }i<\lceil\hat{x}\rceil,\\
      \lceil\hat{x}\rceil-\hat{x},&\text{if }i=\lceil\hat{x}\rceil,\\
      1,&\text{elsewhere}.
    \end{array}
    \right.
    \label{eq:x-dim-overlap}
  \end{equation}

  On the other hand, consider function $g(\cdot)$ in~\Cref{eq:3d-density-accum-corner}. It is clear that we have
  \begin{equation}
    g(i-\hat{x})=\left\{
    \begin{array}{ll}
      \lceil\hat{x}\rceil-\hat{x},&\text{if }i=\lceil\hat{x}-1\rceil,\\
      \hat{x}+1-\lceil\hat{x}\rceil,&\text{if }i=\lceil\hat{x}\rceil,\\
      0,&\text{elsewhere}\\
    \end{array}
    \right.
  \end{equation}
  for any integer $i$. Apply the 1D prefix sum on this function, then it is straightforward to see
  \begin{equation}
    \sum_{i'=0}^{i-1}g(i'-\hat{x})=\frac{\mu([x,d_x]\cap b_x)}{w_b},
    \label{eq:basic-equality}
  \end{equation}
  according to~\Cref{eq:x-dim-overlap}. Now, consider the prefix sum of $\mathcal{A}^{(x,y,z)}$, defined by $\mathcal{P}=\varphi(\mathcal{A}^{(x,y,z)})$. Combining~\Cref{eq:def-prefix-sum-3d,eq:3d-density-accum-corner,eq:x-dim-overlap}, we have
  \begin{equation}
    \begin{aligned}
      \mathcal{P}_{ijk}&=\sum_{i'=0}^{i-1}\sum_{j'=0}^{j-1}\sum_{k'=0}^{k-1}g\left(i'-\hat{x}\right)g\left(j'-\hat{y}\right)g\left(k'-\hat{z}\right)\\
      &=\frac{1}{w_bh_bd_b}\prod_{\mathrm{cyc}}\mu([x,d_x]\cap b_x)
      =\frac{\mu(D^{(x,y,z)}\cap b)}{\mu(b)},\\
    \end{aligned}
  \end{equation}
  and therefore we obtain $\varphi(\mathcal{A}^{(x,y,z)})_b=\frac{\mu(D^{(x,y,z)}\cap b)}{\mu(b)}$.
\end{proof}

Now we are going to complete the proof to~\Cref{thm:basis-prefix-sum-3d} with the help of~\Cref{lem:prefix-sum}.
\begin{proof}
  Denote $\mu_b(\Omega)=\frac{\mu(\Omega\cap b)}{\mu(b)}$ for any mesurable region $\Omega$. According to the \emph{inclusion-exclusion principle}, we have the following relationship
  \begin{equation}
    \mu_b\left(D_v\right)=
    \sum_{\sigma_x,\sigma_y,\sigma_z}
    -\sigma_x\sigma_y\sigma_z\mu_b\left(D^{(x+\sigma_x\frac{w}{2},y+\sigma_y\frac{h}{2},z+\sigma_z\frac{d_z}{4})}\right),
  \end{equation}
  where variables $\sigma_x,\sigma_y,\sigma_z$ are taken over $\{-1,1\}$.
  The prefix sum operator $\varphi$ is linear. Hence, we have
  \begin{equation}
    \begin{aligned}
      \varphi(\mathcal{A}_v)_b&\makebox[1.5em][c]{$=$}\sum_{\sigma_x,\sigma_y,\sigma_z}
      -\sigma_x\sigma_y\sigma_z\varphi\left(\mathcal{A}^{(x+\sigma_x\frac{w}{2},y+\sigma_y\frac{h}{2},z+\sigma_z\frac{d_z}{4})}\right)_b\\
      &\makebox[1.5em][c]{$\mathrel{\overset{(*)}{=}}$}\sum_{\sigma_x,\sigma_y,\sigma_z}-\sigma_x\sigma_y\sigma_z\mu_b\left(D^{(x+\sigma_x\frac{w}{2},y+\sigma_y\frac{h}{2},z+\sigma_z\frac{d_z}{4})}\right)\\
      &\makebox[1.5em][c]{$=$}\mu_b(D_v)=\frac{\mu(D_v\cap b)}{\mu(b)},
    \end{aligned}
  \end{equation}
  where the equation marked with symbol $(*)$ holds according to~\Cref{lem:prefix-sum}. The proof is completed.
\end{proof}

\end{document}